# Release Date Optimization in MRP Using Clearing Functions

Wolfgang Seiringer[a], Klaus Altendorfer[a], Reha Uzsoy[b]

[a]*Department for Production and Operations Management,*
*University of Applied Sciences Upper Austria, A-4400 Steyr, Austria,*
Wolfgang.Seiringer@fh-steyr.at

[b]*Edward P. Fitts Department of Industrial and Systems Engineering, Campus Box 7906*
*North Carolina State University, Raleigh, NC, 27519, USA*
*ruzsoy@ncsu.edu*

## Abstract

This paper integrates a clearing function (CF)-based release planning approach into Material Requirements Planning (MRP) to address its limitations in modeling capacity constraints and dynamic lead times. The proposed optimization model replaces MRP's backward scheduling step while preserving its overall structure. Performance is evaluated through simulation experiments on two flow shop systems that explore a range of demand uncertainties and utilization levels. Computational results show that the proposed approach is capable of yielding significant improvements over the conventional backward scheduling approach, due to its ability to compute planned lead times for individual production orders as opposed to BOM items.

# 1. Introduction

The goal of production planning is to schedule the release of work so that system output aligns with demand. A key challenge is that release decisions must be made before cycle times are known, even though cycle times depend on resource utilization, which is itself affected by these release decisions. In discrete parts manufacturing, cycle times are workload-dependent random variables , making them an endogenous outcome of the planning process rather than an exogenous input (Missbauer and Uzsoy 2022).

Material Requirements Planning (MRP) remains widely adopted in industrial practice (Orlicky 1975; Baker 1993; Vollmann, Berry, and Whybark 1997) and continues to receive significant academic interest. Its core procedure (MRP-I) provides an intuitive, computationally efficient framework for planning material flows through four steps—netting, lot-sizing, backward scheduling, and bill-of-material explosion. Backward scheduling computes planned order releases using deterministic, item-specific planned lead times (PLTs) that do not adjust to changes in workload. Standard descriptions of the basic MRP-I procedure can be found in Hopp and Spearman (2011), Vollmann et al. (2005) and Baker (1993).

A key limitation of MRP-I lies in its simplistic treatment of manufacturing capacity, which is manifested in the use of item-specific exogenous planned lead times as deterministic parameters. While a positive planned lead time for a production resource suggests finite capacity – an infinite capacity resource should, in principle, result in instantaneous production – MRP-I assumes that production resources can deliver material within the planned lead time regardless of their workload, contradicting the dependency between workload and realized cycle times. Several approaches to address this issue are reviewed in Section 2.

This paper explores potential improvements by replacing MRP-I's backward scheduling step with an optimization model using clearing functions (CFs) at each level of the Bill of Material (BOM) tree, without modifying other components of the MRP-I procedure. Lot sizing and order release decisions are made sequentially at each level of the BOM tree as in conventional MRP-I, but release planning decisions are made using the optimization model instead of backward scheduling. Our current approach relies on several admittedly restrictive assumptions: (a) all items sharing a production resource must lie within the same level of the BOM tree, so interactions across BOM levels are not modeled; (b) the integrity of the production lots generated by MRP-I's lot-sizing step is fully preserved, limiting the degree of rescheduling flexibility; and (c) production orders must be released in nondecreasing order of their planned receipt dates, ensuring consistency with preceding MRP-I calculations but constraining the sequencing options available to the CF model. These restrictions define the scope of the proposed integration and clarify that the approach enhances, rather than replaces, core MRP logic.

This very circumscribed application of optimization models clearly cannot yield a global optimal solution to the complex multi-level capacitated lot-sizing problems encountered in discrete parts manufacturing. However, it may lead to improved solutions over those obtained using exogenous PLTs, and also exposes issues that need to be addressed in implementing such enhancements to the MRP-I procedure. The CF approach implemented is quite standard, and does not represent any advancement in that domain. Rather, our contribution lies in exploring the potential for improvement by a naïve implementation of CF-based release planning models within the MRP-I procedure under specific assumptions. In particular, the proposed approach allows PLTs to be computed endogenously for each production order rather than for BOM items as in MRP-I and many of its capacitated extensions. For simplicity of notation, we will refer to the MRP-I procedure simply as “MRP” for the remainder of the paper.

The next section reviews related work on production planning and MRP. Section 3 introduces the CF-based release model with which we replace MRP’s backward scheduling step. We then present simulation experiments on two production systems and conclude with key findings and future research directions.

# 2. Previous related work

Both academic research and industrial practice agree that production planning activities should follow a hierarchical structure, separating material flow planning from production control (de Kok and Fransoo 2003). Production planning is carried out in advance of actual production to coordinate the flow of materials across the different resources in the production system (Missbauer and Uzsoy 2022). This is operationalized through the order release function, which determines when orders are released to the shop floor and transfers control from planning to production, whose objective is to complete each order by its due date.

The time-phased planned order releases generated by MRP fulfill this role. The realized cycle time of a production order—from release to completion—is a random variable influenced by factors such as product mix, process plans, and dispatching rules, but most strongly by resource utilization. Queueing theory, simulation, and empirical studies show that the mean and variance of cycle time increase nonlinearly with utilization (Hopp and Spearman 2011; Curry and Feldman 2011). Since utilization is driven by workload, which, in turn, is determined by order release decisions, realized cycle times are an outcome of order release decisions, not an input. Infinite capacity ought to imply instantaneous production, so the presence of a positive PLT in a production planning model suggests some limit on production capacity in a given time period. Throughout this paper we shall use the term ”cycle time” to denote the random variable representing the realized cycle time of a production order, and ”planned lead time” (PLT) to denote a point estimate derived from the distribution of this random variable for the purposes of order release planning.

Three approaches have been proposed to address the dependency between order release and cycle times. The most widely applied approach, used in MRP and many mathematical models for production planning (de Kok and Fransoo 2003; Missbauer and Uzsoy 2020; Pochet and Wolsey 2006; Voß and Woodruff 2003) and inventory control (Zipkin 2000; Axsäter 2010), relies on deterministic, workload-independent PLTs. Material entering the system at time $t$ is assumed to complete at time $t + L$, where $L$ is a deterministic, externally specified parameter. Early models of this type assumed constant lead times for each item across the entire planning horizon, which is equivalent to assuming a time-stationary cycle time distribution that, in turn, implies a constant average workload in every planning period. Later work has sought to develop time-dependent PLTs, which do not materially increase the complexity of the planning models but pose significant difficulties in estimating a consistent set of PLTs across a planning horizon (Missbauer and Uzsoy 2020, Chapter 6). While this method is intuitive and leads to tractable mathematical models, it reduces the cycle time distribution in a planning period to a single point estimate.

The second approach involves the use of a nonlinear clearing function that relates the workload assigned to a resource in a planning period to its expected output during the period. This approach is adopted in this paper, and will be discussed in detail in Section 2.2.

The third approach uses iterative multi-model methods that alternate between determining order releases based on given PLTs and estimating the resulting cycle times until convergence. These approaches, discussed by Missbauer (2020) and in Chapter 6 of Missbauer and Uzsoy (2020), suffer from inconsistent convergence and high computational effort, especially when cycle time estimation requires multiple simulation replications. We are not aware of any applications of these approaches to multi-level BOMs and therefore do not discuss them further.

## 2.1 Estimation of Planned Lead Times

The problem of estimating PLTs for orders moving through a production system arises in several contexts: the quotation of due dates to customers in make to order environments (Keskinocak and Tayur 2004) and the release of production or replenishment orders in production or inventory systems. Estimates of the remaining time before a particular production lot is completed are also used in shop-floor dispatching rules such as the Apparent Tardiness Cost and COVERT rules (Pinedo 2016) as well as workload-control based order release policies (Missbauer and Uzsoy (2020) Chapter 4). A wide variety of data-driven approaches for predicting the realized cycle time of individual production orders have been proposed (Schneckenreither, Haeussler, and Gerhold 2021). Our discussion focuses on production order release in the MRP environment.

The importance of the PLTs to the effective performance of production planning systems, in particular MRP, was recognized from an early date (St. John 1985; Ho and Lau 1994; Walker and Wysk 1983). Kanet (1986) discusses how the choice of PLTs affects the performance of MRP-based planning systems, and (Graves 2022) provides an excellent perspective on the importance and role of PLTs in production planning.

In production and inventory systems, estimating PLTs is commonly framed as minimizing the expected cost of deviations between deterministic PLTs and realized cycle times, balancing late and early completion costs under known cycle time distributions, with WIP-related costs sometimes included. Weeks (1979) considers a single period static formulation of the problem considering the tradeoff between inventory holding costs due to early completion and tardiness costs incurred by exceeding the PLT. He notes the problem has the mathematical structure of the newsvendor problem (Khouja 1999; Qin et al. 2011) with a critical fractile solution determined by the earliness and tardiness costs. This approach has been extended to a variety of multi-item systems, exploiting the newsvendor structure of the solutions (Gong, de Kok, and Ding 1994; Atan et al. 2016; Jansen et al. 2019). Dolgui and Ould-Louly (2002) consider a multiperiod problem with PLTs taking discrete values, and formulate a Markov chain based on an auxiliary system state space which they use to formulate analytical expressions for the expected earliness and tardiness costs. Ould-Louly and Dolgui (2002) study a special case where multiple items share a common cycle time distribution. Subsequent work extends these ideas to multi-level assembly systems under different lot-sizing policies and service level constraints (Ben-Ammar, Dolgui, and Wu 2018; Hnaien, Dolgui, and Ould Louly 2008; Louly and Dolgui 2008, 2013). Matsuura and Tsubone (1993) take a different approach, seeking an optimal tradeoff between WIP levels and variation in capacity loading across periods. They formulate the problem for a single resource from a queueing perspective, postulating a piecewise constant relationship between the processing time of a production order and its cycle time. PLTs that minimize capacity loading variations for a given average WIP level are computed using a Lagrangian function. Matsuura, Tsubone, and Kanezashi (1996) extend the approach to problems with multiple resources. Several authors (Albey and Uzsoy 2015; Homem-de-Mello, Shapiro, and Spearman 1999; Molinder 1997) use simulation optimization to determine optimal PLTs; this technique is broadly applicable, but is computationally demanding. Albey, Yanikoglu, and Uzsoy (2017) suggest a robust optimization approach with promising results.

These approaches estimate PLTs for BOM items under known demand without capacity constraints; others estimate PLTs for individual orders using capacitated optimization or detailed scheduling. Milne, Mahapatra, and Wang (2015) propose an integer programming approach to compute cost-minimizing planned lead times for MRP in a multi-level BOM with capacity constraints in which planned orders are created in a manner emulating MRP logic. Zijm and Buitenhek (1996) schedule jobs released each period using a heuristic that prevents delaying jobs already in the shop, and use the resulting completion times to compute due dates.

Several authors obtain capacity-aware PLTs by incorporating heuristics into the MRP procedure at each BOM level. Enns and Suwanruji (2004) suggest updating PLT estimates over time using exponential smoothing of prior estimates with observations from the shop floor. Taal and Wortmann (1997) suggest a procedure in which planned order releases are simulated forward in time to identify capacity issues and mechanisms such as adjusting lot sizes, use of alternative routings, iteratively adjusting lead times and simulating, similar to Jodlbauer and Reitner (2012) and Jodlbauer and Strasser (2019). Pandey, Yenradee, and Archariyapruek (2000) propose forward/backward scheduling and order cancellation to ensure feasibility. Kanet and Stößlein (2010) consider capacity before BOM explosion. Tardif and Spearman (1997) developed a method to meet capacity constraints or identify infeasibilities and suggest input changes. Wuttipornpun and Yenradee (2004) propose a Finite Capacity MRP (FCMRP) that reallocates jobs and adjusts timing under lot-for-lot lot sizing, assuming one part per bottleneck machine. Ornek and Cengiz (2006) combine linear programming and MRP, using lot size flexibility, routing, and overtime to ensure feasibility, though setup cost inclusion increases complexity. Wuttipornpun and Yenradee (2007) refine bottleneck scheduling under strict assumptions, later extended by Sukkerd and Wuttipornpun (2016) via a hybrid genetic algorithm and tabu search to reduce computation time. While this type of approach is generally discussed in the context of capacitated MRP, they can also be viewed as approaches to computing PLTs that account for limited capacity. A natural extension is to formulate exact multilevel optimization models that incorporate capacity constraints across the full planning horizon and BOM. While such models are computationally tractable when lot-sizing considerations can be ignored (Tardif and Spearman 1997; Voß and Woodruff 2003; Asmundsson et al. 2009; Asmundsson, Rardin, and Uzsoy 2006), when lot sizing decisions are considered they result in computationally demanding mixed-integer programs (Kimms 1997; Buschkühl et al. 2010).

In contrast to these approaches, our work does not optimize PLTs as planning parameters. We employ the standard MRP procedure to generate planned orders but replace the backward scheduling step with a clearing function-based release planning procedure that considers all items at that level that compete for capacity on a given production resource. This allows release dates to be dynamically adjusted within admissible lead time windows, explicitly incorporating finite capacity. This situates our contribution within the PLT literature in MRP while extending it toward capacity-constrained and stochastic production planning.

## 2.2 Clearing Functions for Production Planning Models

A clearing function (CF) is a surrogate model of a production system that describes the relationship between the expected output of the system, which may be a single machine or a group of machines, in a planning period and the amount of work made available to the system in the planning period. The concept is closely related to flow-density relationships in traffic modeling, which describe traffic flow through a road segment as a function of vehicle density during a given time interval (Carey and Bowers 2012), particularly within the framework of dynamic traffic assignment problems (Peeta and Ziliaskopoulos 2001).

Figure 1 shows several CFs. The “Constant Level” function sets a fixed upper bound on output, assuming instantaneous production at that level for any workload. Graves (1986) proposes a constant proportion CF of the form $Y_t = \alpha W_t$, where output $Y_t$ at time $t$ is a linear function of workload. This function assumes that the facility will be managed to maintain a fixed lead time of $1/\alpha$ at all utilization levels, and may predict infeasible levels of output at high workload levels. The combined CF addresses this by imposing a fixed upper bound on the constant proportion function. The most common (Missbauer and Uzsoy 2020) to use a CF that is concave and monotonically non-decreasing in the workload, shown as the effective CF in the figure, where output increases with workload at a decreasing rate due to congestion. When implemented in mathematical programming models for release planning (Asmundsson et al. 2009; Missbauer and Uzsoy 2020) this yields computationally tractable convex optimization models that can be solved to global optimality using available commercial solvers. Gopalswamy and Uzsoy (2021) have shown that many commonly used CFs of this form can be represented as second-order conic constraints, resulting in significant gains in computational efficiency. A detailed discussion of CFs can be found in Missbauer and Uzsoy (2020).

In CF models, item-level PLTs are not required. The CF specifies resource output, so PLTs vary with workload for each production order. Several approaches have been suggested for estimating CFs, including data-driven approaches (Gopalswamy and Uzsoy 2019; Gopalswamy, Fathi, and Uzsoy 2019; Haeussler, Stampfer, and Missbauer 2020; Kacar and Uzsoy 2015) and queueing models (Karmarkar 1989; Missbauer 2002).

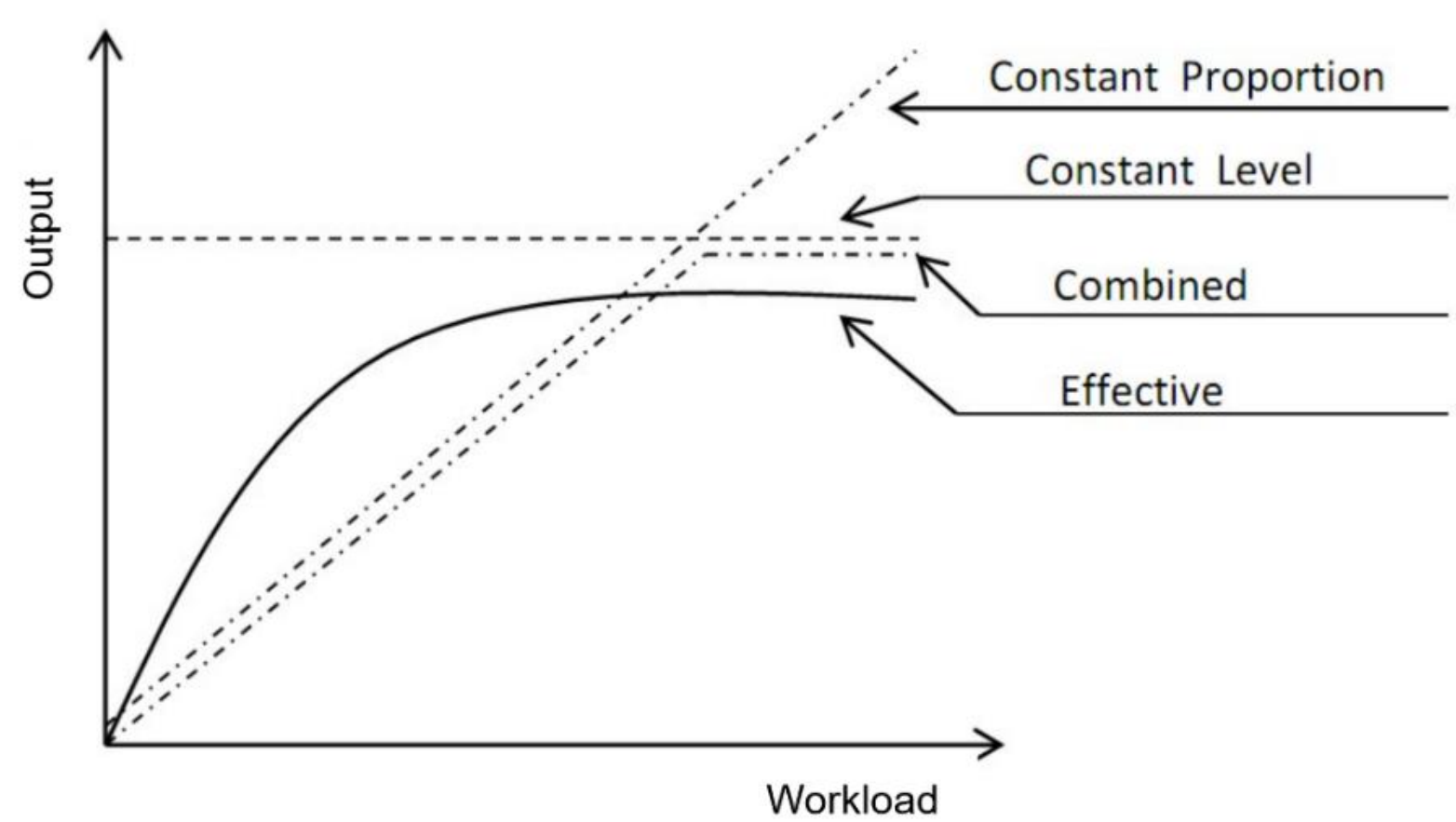

Figure 1: Examples of Clearing Functions, adapted from Karmarkar (1989)

Computational studies (Kacar, Mönch, and Uzsoy 2013; Kacar, Mönch, and Uzsoy 2016; Irdem, Kacar, and Uzsoy 2010; Kacar, Irdem, and Uzsoy 2012; Haeussler, Stampfer, and Missbauer 2020) indicate that CF models often outperform linear programming approaches with fixed PLTs, particularly under time-varying workloads. More recent contributions like Ziarnetzky, Mönch, and Uzsoy (2018) and Ziarnetzky, Mönch, and Uzsoy (2020) apply these models in settings with stochastic demand, combining them with models of forecast evolution.

Early CF models faced challenges in accurately capturing the dynamics of multi-item production systems, which are addressed by the Allocated Clearing Function (ACF) approach developed by Asmundsson, Rardin, and Uzsoy (2006) and later refined by Asmundsson et al. (2009). This paper does not use the ACF model but combines simple CFs with integer sequencing variables to model MRP's backward scheduling step. Aside from these sequencing variables, the CF application is standard; the contribution lies in exploring the benefits of using a CF model for release planning instead of MRP's backward scheduling.

# 3. Model development

The classic MRP procedure executes four steps at each BOM level: netting computes time-phased net requirements, lot sizing determines order quantities and receipt dates, backward scheduling sets release dates using PLTs, and BOM explosion derives component demands (Hopp and Spearman 2011; Orlicky 1975).

In this paper we replace the backward scheduling step by a CF-based optimization model as depicted in Figure 2 while maintaining the other three MRP steps unchanged. We assume that production resources (machines) are shared only within the same BOM level and that lots are released in nondecreasing order of their scheduled receipt dates to remain consistent with lot sizing. The planned order receipts serve as input to a CF-based optimization model that adjusts release dates within admissible lead-time windows to reflect capacity constraints.

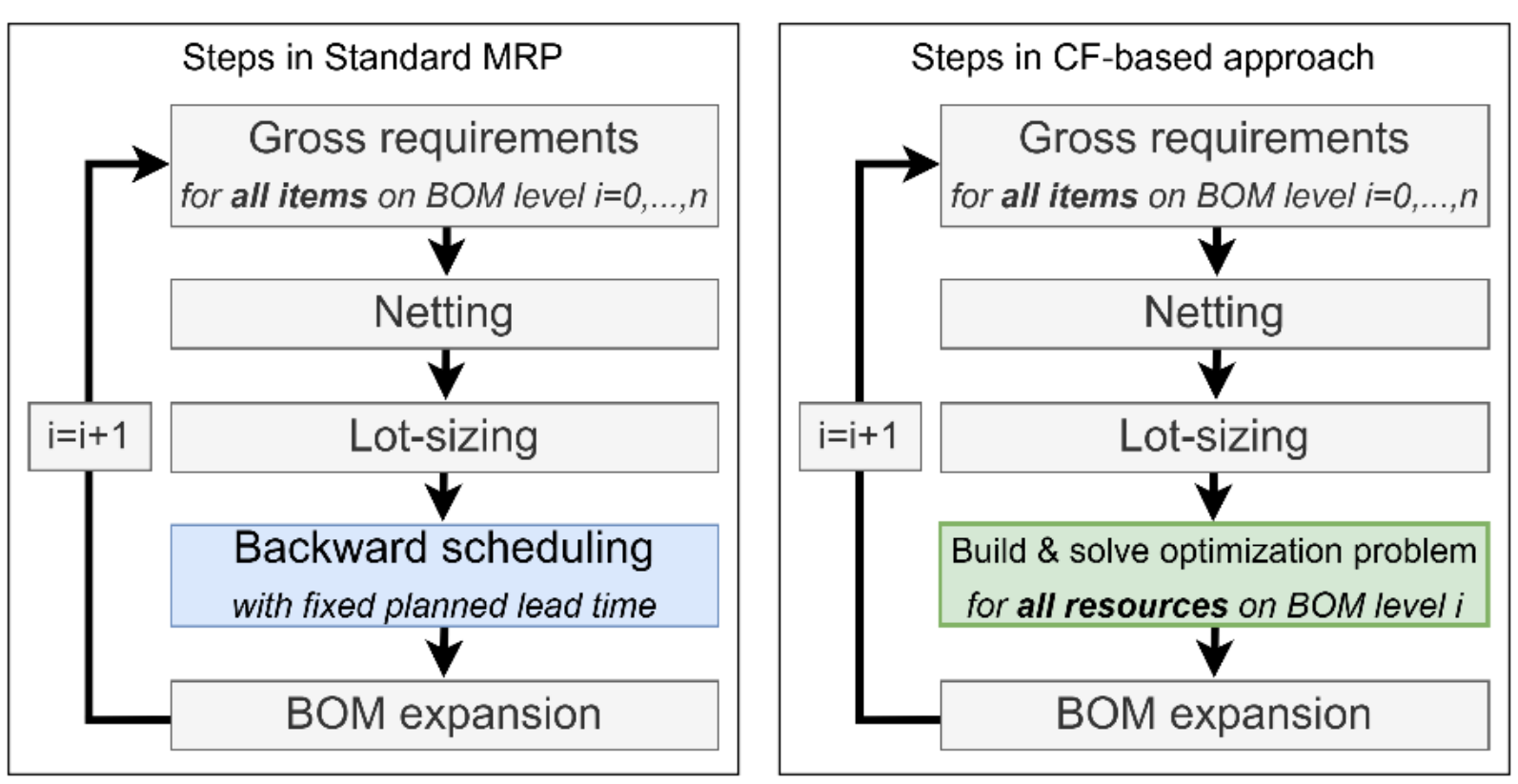

Figure 2: MRP vs. CF release planning in MRP

Figure 3 compares the results of MRP's backward scheduling with a CF formulation for a given BOM level where, for simplicity of exposition, all items have a unit processing time of 20 minutes.

The horizontal axis shows the planning periods with their total capacity loads; each period has 1440 minutes of available capacity.

MRP assumes fixed lead times of 2 periods for each item and computes releases to meet due dates by backward scheduling. The optimization model adjusts release dates within admissible lead-time windows to reflect limited capacity, releasing orders earlier ahead of congested periods with high-capacity requirements and postponing when excess capacity is available. Thus, around periods 4-5 in Figure 3, where capacity loading is heavy, MRP releases Item 102 (Qty 35) and Item 103 (Qty 5) in period 4 for a planned completion in period 6; CF advances these release times to period 3, spreading work out of the congested window. Later in the horizon, excess capacity is available: MRP releases Item 104 (Qty 40) in period 6 for completion in period 8, while CF postpones it to period 7, meeting the due date while reducing WIP. Thus, the optimization model computes a PLT for each production order, as opposed to the item-based PLTs used in MRP, giving it additional flexibility to manage planned order releases to optimize performance. The optimization model is executed at each level of the BOM tree, and production resources are only shared at a given BOM level. This assumption is restrictive, as different BOM levels often share the same equipment. Relaxing it would require simultaneous consideration of multiple BOM levels and remains a topic for future research.

We now formulate the optimization problem for each production resource $r \in E$ used at a given level of the BOM using the notation in Table 1. The model, adapted from Kacar and Uzsoy (2015) and Gopalswamy and Uzsoy (2019), seeks to minimize the sum of inventory, WIP and backlog costs for all items using a piecewise linearized workload-based clearing function (CF) with $C$ segments (constraints 9-12). To simplify notation, we drop the resource index, since each optimization problem is solved for a single resource; thus all model parameters are associated with the specific resource. Assuming production orders are indexed by nondecreasing due dates $d_j$, the model is as follows:

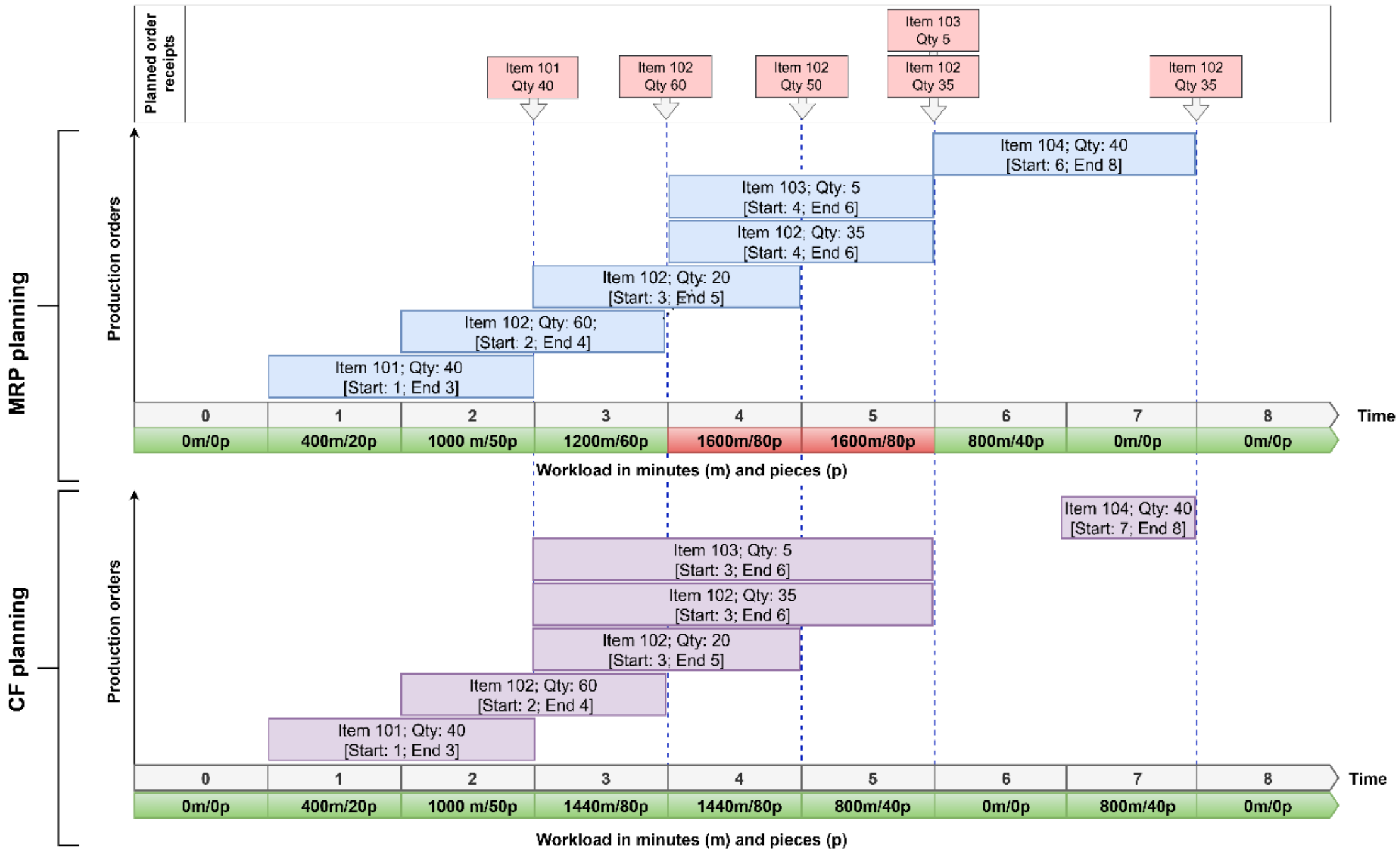

Figure 3: Conceptual comparison of MRP and CF under common planned order receipts

Table 1: Summary of Notation

| **Sets** | |
|---|---|
| $E$ | Set of all production resources, with elements $r \in E$. |
| $G$ | set of all items at the current BOM, with elements $g \in G$. |
| $T$ | set of planning periods, with elements $t \in T = \{1, \dots, T\}$ |
| $J$ | set of production orders, with elements $j \in J$. |
| **Parameters** | |
| $X_{gj}$ | Lot size of production order $j$ of item $g$ due at the end of period $d_j$ as determined by the MRP lot-sizing step. |
| $h_g^I$ | Unit finished inventory holding cost for item $g$. |
| $h_g^W$ | Unit WIP holding cost for item $g$. |
| $b_g$ | Unit backlog costs for item $g$. |
| $o$ | Unit cost of external capacity, set to $o = 1000 \sum_{g \in G} h_{Ig}$ to create a high penalty. |
| $d_j$ | Planned delivery date of production order $j$, i.e. $X_{gjt} = 0 \quad \forall t \neq d_j$ . |
| $l$ | Maximum number of periods before $d_j$ an order may be released. |
| $p_g$ | Unit processing time of item $g$. |
| $s_g$ | Sequence-independent setup time for item $g$. |
| $\Delta$ | Maximum processing capacity per period (time units). |
| $\alpha_x, \beta_x$ | Clearing function parameters. |

| Decision Variables | |
|---|---|
| $RI_{gjt}$ | Binary variable equal to 1 if order $j$ of item $g$ is released in period $t$, and 0 otherwise. |
| $I_{gt}$ | Completed units of item $g$ available at the end of period $t$. |
| $W_{gt}$ | Released units of item $g$ still in process at the end of period $t$. |
| $B_{gt}$ | Backlog units of item $g$ at the end of period $t$. |
| $C_t$ | External capacity used in period $t$ (time units). |
| $P_{gt}$ | Amount of item $g$ produced in period $t$. |
| $R_{gt}$ | Amount of item $g$ released in period $t$. |
| $L_t$ | System load in period $t$ (time units). |
| $PL_t$ | Processed load in period $t$ (time units). |

The objective function can be stated as follows:

$$Min \sum_{g\in G}\left[\sum_{t=1}^{T}[h_g^I I_{gt} + h_g^W W_{gt} + b_{gt}B_{gt} + C_t o]\right] \tag{1}$$

subject to:

$$R_{gt} = RI_{gjt}X_{gj} \quad \forall g \in G, \forall j \in J, \forall t \in T \tag{2}$$

$$\sum_{t=\,d_j-l}^{d_j-1} RI_{gjt} = 1, \forall g \in G, \forall j \in J \tag{3}$$

$$W_{gt} = W_{g,t-1} + R_{gt} - P_{g,t-1} \quad \forall g \in G \; and \; \forall t \in T \tag{4}$$

$$I_{gt} = I_{g,t-1} - X_{gt} + P_{g,t-1} - B_{g,t-1} + B_{gt} \quad \forall g \in G \; and \; \forall \, t \in T \tag{5}$$

$$\sum_{t=1}^{T} tRI_{gjt} \le \sum_{t=1}^{T} tRI_{g,j+1,t} \quad \forall j \in J, \forall g \in G \tag{6}$$

$$PL_t \le \; \alpha_c L_t + \; \beta_{c,} \, c = 1, \dots, C \tag{7}$$

$$I_{gt}, W_{gt}, P_{gt}, B_{gt}, L_t \ge 0, RI_{gt} \in \{0,1\} \tag{8}$$

The objective function (1) minimizes the sum of finished inventory holding, WIP holding, backordering and external capacity costs. The external capacity option is included at a high cost to identify infeasible instances due to limited capacity. Constraints (2) computes the release quantity related to production order $g$ in period $t$ in terms of the discrete variables $RI_{gt}$ specifying the release date of the production order. Constraints (3) ensure that each order $g$ is released exactly once during the planning horizon $t$. These constraints are required to preserve the integrity of the production lots determined by the MRP lot-sizing procedure. Constraints (4) and (5) are material balance equations for finished goods and WIP inventory, respectively, of each item. Constraints (2) are shown explicitly for clarity of exposition; in the implemented model they can be substituted directly into (4), yielding

$$W_{gt} = W_{g,t-1} + RI_{gjt}X_{gj} - P_{g,t-1} \quad \forall g \in G \; and \; \forall t \in T. \tag{9}$$

Constraints (6) ensure that no production order with a later scheduled receipt date is released before another with another with an earlier date. This clearly limits the ability of the optimization model to obtain cost-minimizing release plans, but is necessary to maintain the consistency of the release plans with the preceding lot-sizing calculations.

The limited capacity of the production resource is represented in constraints (7) using a piecewise linearized CF that relates the system load in a planning period $t$ to the output or processed load of the resource in that period, with all quantities measured in units of processing time. The system load $L_t$ denotes the total amount of work, in time units, made available for processing during the planning period and can be calculated as:

$$L_t = \sum_g W_{gt}p_g + s_g \sum_j RI_{gjt} \tag{10}$$

where $p_g$ denotes the unit processing time for item $t$ and $s_g$ the setup time for a production order of item $g$. Similarly, the processed load $PL_t$ at the resource in period $t$, in units of time, represents the total time consumed to produce $P_{gt}$ units of all items $g$ and is given by:

$$PL_t = \sum_g P_{gt}p_g + s_g \sum_j RI_{gjt} \tag{11}$$

Most CFs used in the literature are assumed to be concave and monotonically non-decreasing, as in the Effective Clearing Function in Figure 1. It is well known (Missbauer and Uzsoy 2020) that any such function $f(L_t)$ can be represented to an arbitrary degree of accuracy by the convex hull of a set of affine functions of the form

$$f_c(L_t) = \alpha_c L_t + \beta_c, c = 1, \dots, C. \tag{12}$$

To capture the concavity of the underlying CF we assume that $\alpha_1 = 1 \geq \alpha_2 \geq \cdots \geq \alpha_C$ and $\beta_1 = 0 \leq \beta_2 \leq \cdots \leq \beta_C = \Delta$. In this paper we use two simple CFs to examine the impact of the optimization model on the performance of the overall MRP procedure. The first of these is a simple two-segment CF consisting of

$$\sum_g P_{gt}p_g + s_g \sum_j RI_{gjt} \le \sum_g W_{gt}p_g + s_g \sum_j RI_{gjt}$$

$$\sum_g P_{gt}p_g + s_g \sum_j RI_{gjt} \le \Delta \tag{13}$$

The first segment ensures that processed load cannot exceed system load, corresponding to a linear segment with $\alpha_1 = 1$ and $\beta_1 = 0$, while the second represents the maximum processing time available in a time period ($\alpha_2 = 0$ and $\beta_2 = \Delta$). This CF corresponds to a simple capacity limitation corresponding to the Best-Case analysis of Hopp and Spearman (2011), which does not consider uncertainty in processing times. We also consider the addition of an intermediate segment representing loss of output due to congestion, yielding the three segments

$$\sum_g P_{gt}p_g + s_g \sum_j RI_{gjt} \le \sum_g W_{gt}p_g + s_g \sum_j RI_{gjt}$$

$$\sum_g P_{gt}p_g + s_g \sum_j RI_{gjt} \le \alpha_2 \left( \sum_g W_{gt}p_g + s_g \sum_j RI_{gjt} \right) + \beta_2 \tag{14}$$

$$\sum_g P_{gt}p_g + s_g \sum_j RI_{gjt} \le \Delta$$

where the values of $\alpha_2$ and $\beta_2$ are varied according to our experimental design.

This model differs from the Allocated Clearing Function formulation of (Asmundsson, Rardin, and Uzsoy 2006; Asmundsson et al. 2009) in several ways. Unlike that model, which uses continuous decision variables and does not consider lot-sizing, our model preserves the integrity of production lots computed by MRP, requiring the use of the integer variables $RI_{gt}$. The use of allocation variables to ensure consistency of output and input mixes is unnecessary since constraints (6) ensure that lots are released and processed in order of their due dates with no overtaking, and lot integrity is preserved by (2) and (3). These latter restrictions limit the quality of the solutions that can be obtained from the optimization problem, only allowing it to adjust planned order release times within a limited time window. Thus the model generates PLTs for individual production lots as opposed to the item-level PLT estimates common in MRP.

# 4. Computational Experiments

This section compares standard MRP with CF-based planning to assess the benefits of using the CF approach in two production systems. Both systems exhibit diverging BOMs where each item may be required by several items at the next higher level of the BOM tree. We use the same functional form of the CF for all machines. The arguments of the CF (the system and processed load) are calculated for each machine based on the processing and setup times of the items processed on that machine; hence the CFs differ across machines. For both systems, simulations are run over 200 periods (including 40 warm-up periods), with 10 independent replications per parameterization. Statistical significance of results is assessed at the 95% and 99% confidence levels - marked by * and **, respectively - in the result tables. Significance is always computed based on the comparison between MRP and CF under identical parameter settings. Enumeration is used to systematically evaluate planning parameters under controlled, reproducible conditions. Despite its higher computational cost, this exhaustive approach ensures fair and accurate comparisons across all tested combinations. Uncertainty is introduced through customer demand and lognormally distributed setup times (CV=0.2), while processing times are deterministic and differ by item. The log-normal distribution was chosen for its positive skewness and suitability for nonnegative random variables that can attain high values (Crow and Shimizu 2018; Wehrspohn and Ernst 2022).

In the simulation, production orders enter finished goods inventory only after completion of the entire lot. Optimization to proven optimality is performed in each of the 200 planning periods without time limits or MIP gaps. Released production orders are dispatched according to the earliest due date (EDD) rule. This interaction is implemented using the CPLEX Java API, and directly translates optimized release periods into production start dates. The simulation model was implemented in AnyLogic as a fully parameterizable multi-item, multi-stage framework, allowing configuration of forecast settings, production system parameters, and the planning method. Model tuning and the identification of parameter ranges were based on preliminary studies. Simulation experiments ran on a shared pool of 20 computers (up to 40 concurrent instances in total, with maximum of 2 instances per computer). MRP and CF jobs were dispatched to the same pool. A central PostgreSQL server managed experiment parameterization and stored replication results, while an SQLite database provided the production-system master data.

To model demand uncertainty for end items, we use a simplified additive Martingale Model of Forecast Evolution (MMFE) (Heath and Jackson 1994; Güllü 1996; Norouzi and Uzsoy 2014; Altendorfer and Felberbauer 2023). Demand forecasts for finished products are provided for a forecast window consisting of the $H$ periods prior to demand realization and are updated until the final demand realization. The demand forecast $D_{gd\gamma}$ for an end item $g \in G$ with due date $d$ computed $\gamma$ periods before its due date represents the gross requirements to be processed by the netting and lot-sizing steps to obtain the net requirements (dependent demand) $X_{gjt}$ that are the input for the CF-based optimization, and is calculated as follows:

$$D_{gd\gamma} = x_g \quad for\ \gamma > H \tag{15}$$

$$D_{gd\gamma} = x_g + \varepsilon_{gd\gamma} \quad for\ \gamma = H \tag{16}$$

$$D_{gd\gamma} = D_{g,d,\gamma+1} + \varepsilon_{gd\gamma} \quad for\ \gamma < H \tag{17}$$

where $H$ denotes the length of the forecast window and $x_g$ the long-term demand forecast for end item $g$, generally set to the mean demand. $\varepsilon_{gd\gamma}$ is the forecast update term for end item $g$ and due date $d$ obtained $\gamma$ periods before delivery. For $\gamma \leq H$ $\varepsilon_{gd\gamma}$ follows an independent truncated normal distribution with mean 0 and a standard deviation $\sigma_{\varepsilon}^{\gamma}$. We use a truncated distribution to avoid the possibility of negative demand. Setting $\varepsilon_{gd1} = 0$ yields the realized demand with no further updates in subsequent planning periods.

In our experiments we consider two updates with $\gamma = 10$ for the first update and $\gamma = 1$ for the second, which applies a final adjustment one period before delivery. Thus 10 periods before delivery we have $\varepsilon_{g,d,10} \sim N(0, \varphi x_g)$, and one period before delivery we have $\varepsilon_{g,d,10} \sim N(0, \psi x_g)$, with all other updates set to 0. Thus after the update at $\gamma = 10$, the forecast remains constant until the $\psi$ update at $\gamma = 1$, after which it remains constant until the due date. For the perfect forecast case we have $\varphi > 0$ and $\psi = 0$, while for the imperfect forecast case we have $\varphi > 0$ and $\psi > 0$. The values of $\varphi$ and $\psi$ examined in our experiments are summarized in Table 2. An illustrative example of the forecast settings can be found in the Appendix.

Table 2: Experimental Design

| Parameter | Values |
|---|---|
| **Forecast setting** | *Perfect forecast*:<br>• $\varphi \in \{0.25, 0.5, 0.75, 1.0, 1.25, 1.5\},\ \psi = 0.$<br>*Imperfect forecast*:<br>• $\varphi \in \{0.25, 0.75\},\ \psi \in \{0.05, 0.1, 0.25\}$ |
| **Average Utilization** | 80% ($x_g$=41.18), 85% ($x_g$=44.12), 90% ($x_g$=47.06) |
| **Lot-sizing Policy** | FOP 1, FOP 2 and FOP 3 |
| **Safety Stock** | 0, 0.2, 0.4, 0.6, 0.8, 1 and 1.2 |
| **Planned Lead Time** | 1, 2, 3 periods and CF based release planning |
| **Release Planning** | 1) Backward scheduling<br>2) Ideal CF with two segments<br>3) CF with three segments (80% high, 60% medium, 40% low, 20% very low) |

# 5. Results for Simple Production System (PS1)

We begin by comparing standard MRP and CF-based release planning using the simple production system PS1, illustrated in Figure 4. PS1 has three BOM levels: end items 100-103 (produced on machines M101 and M102), components 200-201 (machine M201), and raw materials 300-301 (available instantaneously). Each end item requires one component, and each component one raw material. To ensure comparability across utilization levels, the setup and processing times are held constant, while the long-term demand forecasts $x_g$ for each item $g$ are adjusted equally across all items to achieve the desired average utilization levels given in Table 2. This results in forecast levels of $x_g$=41.18, $x_g$=44.12, and $x_g$=47.06 units for target utilizations of 80%, 85%, and 90%. For each utilization level, all end items have equal mean demand to represent a balanced product mix, while demand and forecast uncertainty create variation in realized demand.

Each machine (M101, M102, M201) has a capacity of $\Delta = 1{,}440$ minutes per period. In the 90% utilization case, 1,296 minutes are used – 1,152 for processing and 144 (10%) for setup– allowing two items per machine and period, each with a 72-minute setup (one lot per item, i.e., FOP 1 policy). For the 85% and 80% cases, we keep the setup time fixed at 144 minutes and reduce only the demand per period, using 1,224 minutes (1,080 processing + 144 setup) and 1,152 minutes (1,008 processing + 144 setup), respectively.

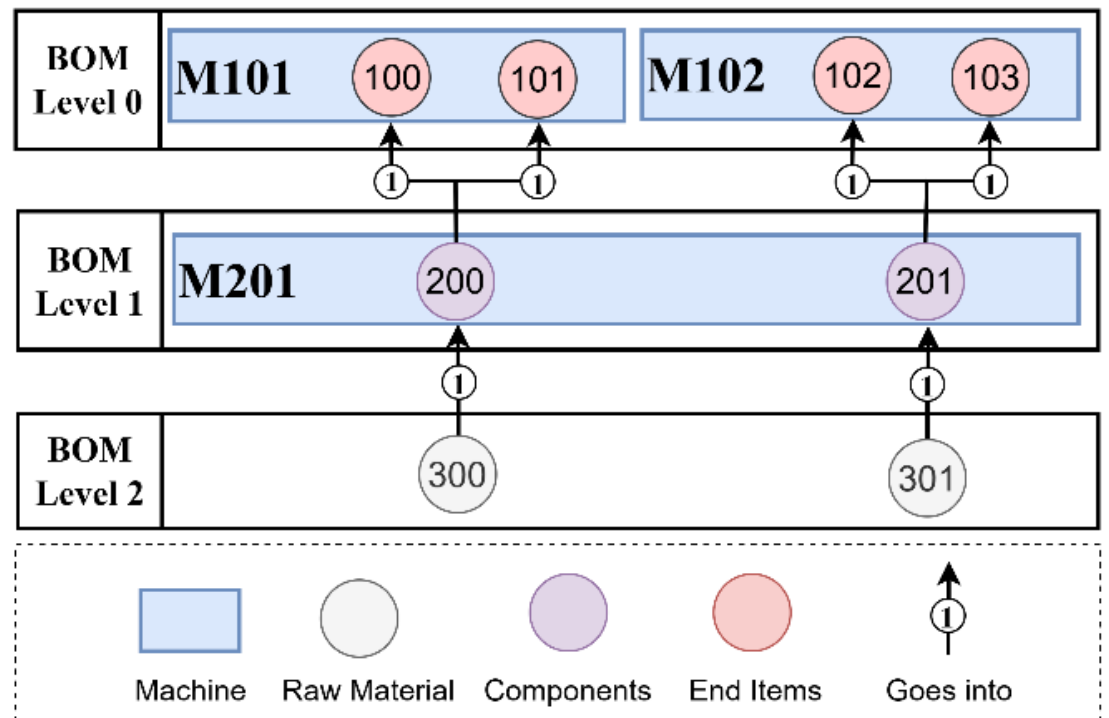

Figure 4: Structure of production system 1 (PS1)

The Fixed Order Period (FOP) policy with $q$ =1, 2, and 3 was used to evaluate its effect on order timing, inventory and tardiness. FOP aggregates demand over $q$ periods to set order receipts and was selected for its simplicity and suitability for controlled comparisons. Safety stock ranged from 0 to 1.2 times average demand (in increments of 0.2) to assess inventory-tardiness trade-offs, and influences the optimization through the netting step. In our experiments, MRP uses fixed PLTs (set to 1, 2 or 3 periods for all items), while the CF-based approach applies the same functional form of the CF across items. All planning parameters, including PLTs and lot sizing, were identical for all end items and components. We first compare CF and MRP performance under different utilization levels and forecast uncertainties, then test three-segment CFs representing different capacity levels.

We present the minimum overall costs for each forecast setting among all the tested parameterizations, expressed in Cost Units (CU). The overall costs are calculated by averaging the sum of Finished Goods Inventory (FGI) for end items and components, Work in Progress (WIP) for end items and components, and tardiness costs in periods (days) over 10 simulation replications. Cost parameters are summarized in Table 3. The tardiness costs are set to 38, based on the newsvendor model targeting a 95% service level. With unit FGI holding cost of 2, the uncapacitated newsvendor model yields $c_u = 38$, corresponding to a 5% stockout probability (Axsäter 2015).

Table 3: Cost settings

| **Cost Category** | **Costing Factor (CU)** | **Description** |
|---|---|---|
| End Items - FGI | 2 | Unit end item finished goods inventory holding cost |
| End Items - WIP | 1 | Unit end item work in progress holding cost. |
| Components - FGI | 1 | Unit component inventory holding costs. |
| Components - WIP | 1 | Unit component work in progress holding cost |
| Tardiness - End Items | 38 | Unit tardiness cost for end items |

## 5.1. Results for PS 1 under Perfect Forecasts

Figure 5 compares MRP and Ideal CF in terms of overall costs at different levels of demand uncertainty and average utilization. Table 4 summarizes service levels, planning parameters, and cost percentages of WIP, inventory and tardiness costs related to overall costs as well as the cost-minimizing lot sizing and safety stock policies. The x-axis shows the forecast settings $\varphi$ and the y-axis the minimum total costs. Data labels indicate the Average Planned Flow Time (APF), defined as the planned end date minus the planned start date. Although MRP uses fixed PLTs, demand updates within the forecast window can cause orders to be expedited or delayed by the planning run in the next period, resulting in different planned flow times across runs. In contrast, CF adjusts releases dynamically, leading to varying APFs across planning runs.

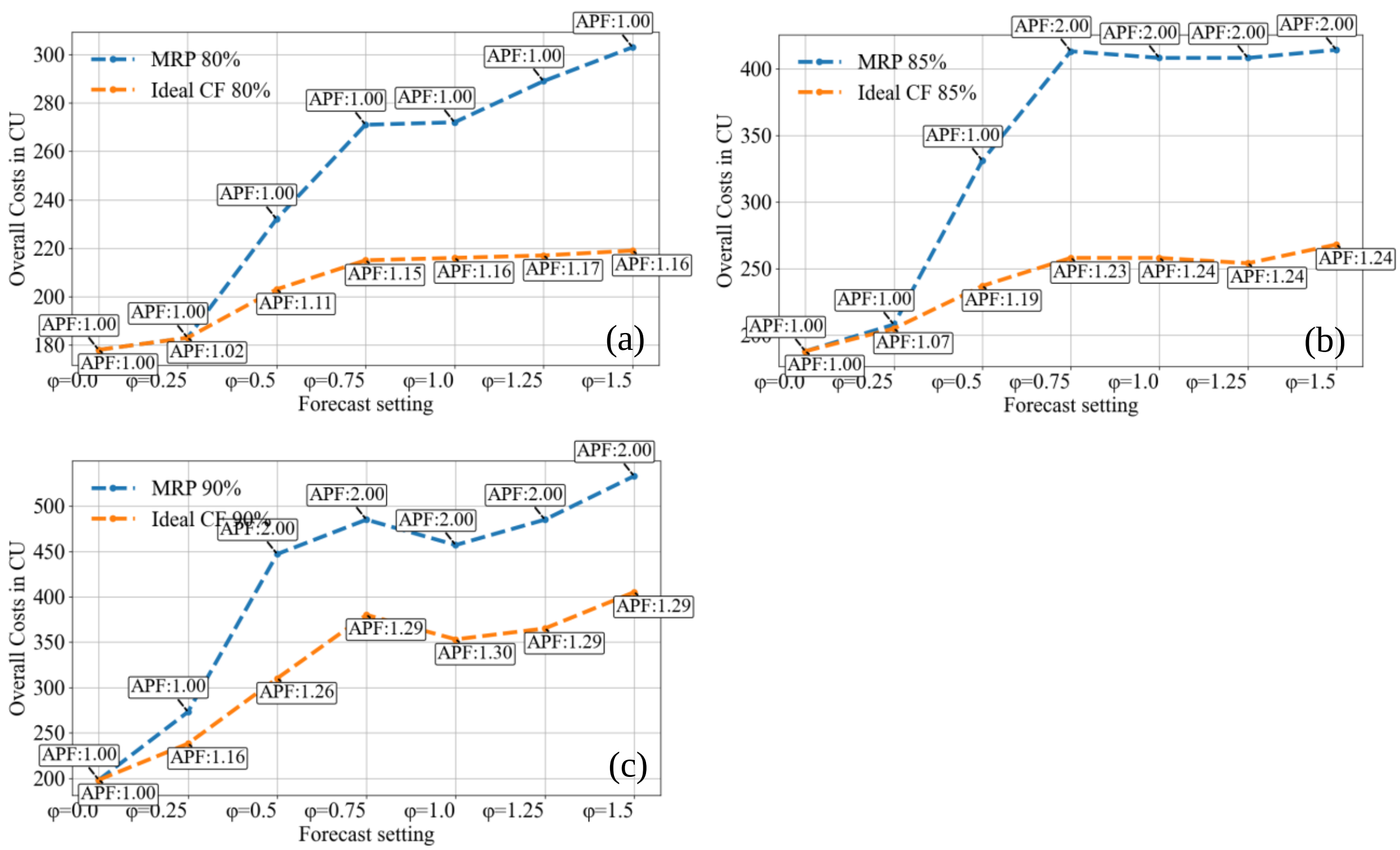

Figure 5: Minimum overall costs for PS1 under perfect forecast information at (a) 80%, (b) 85%, and (c) 90% utilization

At all utilization levels the cost-minimizing lot sizing policy is FOP1, lot-for-lot, while no safety stock is held by either planning model. It is also interesting that at all three utilization levels both procedures appear to produce exactly the same plan for $\varphi = 0$, corresponding to no demand variability. This implies that both procedures release production orders in the period immediately before they are due. As demand variability increases, the differences between the procedures become more defined.

At 80% utilization (Figure 5 (a)), the best performing PLT value is one period, while the APF for CF is between 1 and 2 periods. The higher costs for MRP at $\varphi \geq 0.5$ indicate that orders are released too late. Thus WIP and inventory holding costs remain low, but tardiness costs increase and service level declines, as shown in Table 4(a). CF, by releasing production lots ahead of periods with high demand, keeps costs low and service levels high (≥ 98%), though incurring higher inventory costs. CF effectively assigns a different PLT to each production lot, in contrast to MRP where the PLT parameter is common to all production lots of that item.

At 85% utilization (Figure 5 (b)), the same phenomenon is observed: at low demand variability the best performing PLT for MRP is 1, increasing to 2 for higher levels of demand variability ($\varphi \geq$ 0.75). The CF APF never reaches 2 even for high demand variability but always exceeds 1, suggesting that CF is releasing most, but not all, production orders more than one period ahead of their due date. MRP does not have this flexibility; with PLT = 2, it must release all lots 2 periods ahead of due date, leading to increased inventory levels over CF. At $\varphi = 0.5$, MRP releases lots too late, incurring high tardiness costs that outweigh reduced inventory holding costs. For $\varphi \geq 0.75$, low tardiness costs for MRP are outweighed by high inventory holding costs due to a PLT of 2. At 90% utilization (Figure 5 (c)) the behavior of the two systems is similar to that for 85%. The flexibility of CF to assign different lead times to individual orders appears to be the fundamental reason for CF's improved performance. Under high demand variability, MRP must increase the PLT to 2 periods to avoid high tardiness costs, but CF can absorb demand variations by releasing some orders ahead of periods where demand exceeds available capacity.

These results illustrate the potential improvements that the CF approach can yield. PLTs in MRP are estimated for each item, not individual production lots, and are set to an integer number of planning periods (although a number of continuous-time MRP procedures have been suggested (Hackman and Leachman 1989)). Our findings are consistent with previous work in that workload-dependent, order-specific lead times outperform fixed item PLTs under congestion (Zijm and Buitenhek 1996; Asmundsson et al. 2009; Kacar, Mönch, and Uzsoy 2016). We now examine the performance of the two planning methods under the imperfect forecast setting which introduces a second forecast update one period before the due date.

Table 4: Minimum overall costs (PS1) under perfect forecast information for (a) 80%, (b) 85%, and (c) 90% Utilization

| (a) 80% Utilization | Forecast setting | | | | | | | | | | | | | |
|---|---|---|---|---|---|---|---|---|---|---|---|---|---|---|
| | $\varphi$=0 | | $\varphi$=0.25 | | $\varphi$=0.5 | | $\varphi$=0.75 | | $\varphi$=1.0 | | $\varphi$=1.25 | | $\varphi$=1.5 | |
| **Planning Type** | **MRP** | **CF** | **MRP** | **CF** | **MRP** | **CF** | **MRP** | **CF** | **MRP** | **CF** | **MRP** | **CF** | **MRP** | **CF** |
| **Service level** | 100% | 100% | 96% | 99% | 87% | 99% | 82% | 98% | 82% | 98% | 81% | 99% | 80% | 98% |
| **Safety stock** | 0 | 0 | 0 | 0 | 0 | 0 | 0 | 0 | 0 | 0 | 0 | 0 | 0 | 0 |
| **FOP** | 1 | 1 | 1 | 1 | 1 | 1 | 1 | 1 | 1 | 1 | 1 | 1 | 1 | 1 |
| **PLT** | 1 | - | 1 | - | 1 | - | 1 | - | 1 | - | 1 | - | 1 | - |
| **Overallcosts** | 178 | 178 | 183 | 183 | 232 | 203** | 274 | 216** | 272 | 216** | 289 | 217** | 304 | 219** |
| **Inventory sales** | 35% | 35% | 33% | 34% | 22% | 33% | 17% | 34% | 17% | 33% | 15% | 34% | 14% | 34% |
| **WIP sales** | 18% | 23% | 24% | 25% | 21% | 26% | 20% | 25% | 20% | 26% | 19% | 26% | 9% | 26% |
| **Inventory components** | 23% | 18% | 17% | 17% | 12% | 17% | 10% | 16% | 10% | 16% | 10% | 16% | 18% | 16% |
| **WIP components** | 23% | 23% | 23% | 24% | 19% | 24% | 17% | 24% | 17% | 24% | 16% | 24% | 15% | 23% |
| **Tardiness** | 1% | 1% | 3% | 0% | 26% | 0% | 36% | 1% | 36% | 1% | 40% | 0% | 44% | 1% |

| (b) 85% Utilization | Forecast setting | | | | | | | | | | | | | |
|---|---|---|---|---|---|---|---|---|---|---|---|---|---|---|
| | $\varphi$=0 | | $\varphi$=0.25 | | $\varphi$=0.5 | | $\varphi$=0.75 | | $\varphi$=1.0 | | $\varphi$=1.25 | | $\varphi$=1.5 | |
| **Planning Type** | **MRP** | **CF** | **MRP** | **CF** | **MRP** | **CF** | **MRP** | **CF** | **MRP** | **CF** | **MRP** | **CF** | **MRP** | **CF** |
| **Service level** | 100% | 100% | 92% | 98% | 78% | 98% | 99% | 97% | 100% | 97% | 99% | 97% | 98% | 96% |
| **Safety stock** | 0 | 0 | 0 | 0 | 0 | 0 | 0 | 0 | 0 | 0 | 0 | 0 | 0 | 0 |
| **FOP** | 1 | 1 | 1 | 1 | 1 | 1 | 1 | 1 | 1 | 1 | 1 | 1 | 1 | 1 |
| **PLT** | 1 | - | 1 | - | 1 | - | 2 | - | 2 | - | 2 | - | 2 | - |
| **Overallcosts** | 188 | 188 | 208 | 205 | 336 | 237** | 414 | 265** | 408 | 258** | 408 | 254** | 422 | 269** |
| **Inventory sales** | 34% | 34% | 28% | 32% | 14% | 32% | 44% | 31% | 45% | 31% | 43% | 33% | 41% | 30% |
| **WIP sales** | 25% | 25% | 25% | 26% | 18% | 26% | 16% | 25% | 16% | 26% | 17% | 26% | 16% | 25% |
| **Inventory components** | 16% | 16% | 14% | 16% | 8% | 15% | 25% | 14% | 25% | 15% | 25% | 15% | 24% | 14% |
| **WIP components** | 25% | 25% | 23% | 26% | 15% | 25% | 13% | 23% | 13% | 23% | 13% | 24% | 13% | 22% |
| **Tardiness** | 0% | 0% | 10% | 0% | 45% | 2% | 2% | 7% | 1% | 5% | 2% | 2% | 6% | 9% |

| (c) 90% Utilization | Forecast setting | | | | | | | | | | | | | |
|---|---|---|---|---|---|---|---|---|---|---|---|---|---|---|
| | $\varphi$=0 | | $\varphi$=0.25 | | $\varphi$=0.5 | | $\varphi$=0.75 | | $\varphi$=1.0 | | $\varphi$=1.25 | | $\varphi$=1.5 | |
| **Planning Type** | **MRP** | **CF** | **MRP** | **CF** | **MRP** | **CF** | **MRP** | **CF** | **MRP** | **CF** | **MRP** | **CF** | **MRP** | **CF** |
| **Service level** | 100% | 100% | 82% | 97% | 98% | 94% | 95% | 92% | 96% | 92% | 94% | 91% | 93% | 91% |
| **Safety stock** | 0 | 0 | 0 | 0 | 0 | 0 | 0 | 0 | 0 | 0 | 0 | 0 | 0 | 0 |
| **FOP** | 1 | 1 | 1 | 1 | 1 | 1 | 1 | 1 | 1 | 1 | 1 | 1 | 1 | 1 |
| **PLT** | 1 | - | 1 | - | 2 | - | 2 | - | 2 | - | 2 | - | 2 | - |
| **Overallcosts** | 198 | 198 | 274 | 238** | 458 | 331 | 506 | 387 | 457 | 353** | 485 | 365** | 538 | 405 |
| **Inventory sales** | 30% | 30% | 19% | 29% | 40% | 24% | 33% | 22% | 36% | 24% | 33% | 24% | 29% | 21% |
| **WIP sales** | 27% | 27% | 22% | 27% | 17% | 22% | 17% | 20% | 19% | 22% | 19% | 22% | 17% | 20% |
| **Inventory components** | 16% | 16% | 11% | 15% | 23% | 11% | 20% | 10% | 22% | 11% | 20% | 11% | 18% | 10% |
| **WIP components** | 27% | 27% | 20% | 27% | 14% | 21% | 14% | 19% | 15% | 20% | 14% | 19% | 13% | 18% |
| **Tardiness** | 0% | 0% | 28% | 2% | 6% | 22% | 16% | 29% | 8% | 23% | 14% | 24% | 23% | 31% |

## 5.2. Results for PS1 under Imperfect Forecasts

In this section we consider the second forecast update one period before delivery ($\gamma$=1). At this point, the demand changes according to the $\psi$ values listed in Table 2, bringing substantial additional uncertainty. While SS=0 and FOP=1 minimize costs for MRP under perfect forecasts (Table 4**)**, this no longer holds under imperfect forecasts. Figure 6 (a) and Figure 6 (b) show CF's cost behavior at 80% utilization with $\varphi$ = 0.75 and $\psi$ = 0.25. Figure 6 (a) reveals a convex cost curve over safety stock values, with minimum at SS=0.4. Figure 6 (b) shows that while larger FOPs reduce tardiness by releasing orders for several periods together, they increase inventory costs, with the lowest overall cost (400 CU) at FOP=1. Similar insights are shown in Rosman et al. (2024), where simulating safety-stock targets under forecast volatility exposes non-monotonic cost–service behavior. Likewise, Forel and Grunow (2023) show that larger release batches (higher FOP) reduce tardiness but increase inventory, creating an implicit safety lead time.

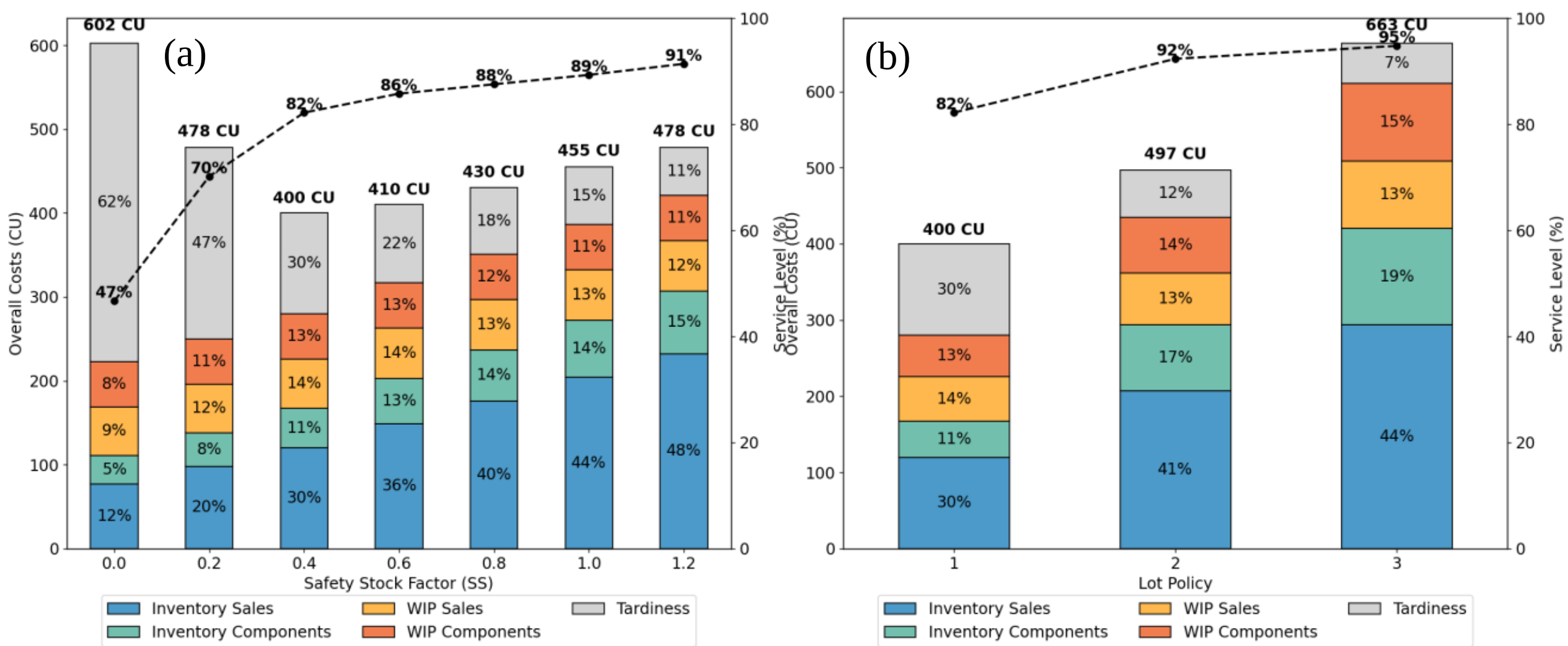

Figure 6: Cost behavior for 80 % utilization with ideal CF and imperfect forecast ($\varphi$ = 0.75, $\psi$ = 0.25)

Figure 7 and Table 5 compare MRP and CF with two demand updates ($\varphi, \psi$) on service levels, planning parameters, and cost components. At 80% utilization (Figure 7 (a), Table 5 (a)), both procedures yield similar total cost for $\varphi$=0.25, with a slight advantage to CF. The APF and inventory costs for both procedures are very similar, although MRP consistently holds higher finished inventory of end items (sales inventory) and its safety stock never exceeds that for CF. For $\varphi$=0.75, CF shows a larger cost advantage. For $\psi$=0.05 and 0.1, MRP retains PLT = 1, but adds safety stock at the higher uncertainty level. For $\psi$=0.25 MRP is forced to increase PLT to 2, obtaining substantially lower tardiness costs than CF with higher inventory levels. At 85% utilization, the same behavior persists, with MRP forced to increase PLT to 2 for all demand scenarios with $\varphi$=0.75. For 90% utilization, both procedures yield similar total costs except for the first demand scenario where CF has a lower cost. As demand uncertainty increases, CF increases the release lot size and safety stock, while the high PLT allows MRP to make these changes at higher uncertainty levels. The APF for CF is consistently lower than that for MRP, suggesting more frequent release of urgent orders due immediately than MRP. Because APF averages integer lead times, a lower APF under CF means more very short lead times (0–1 period). When the second forecast update creates a net requirement in the delivery period, we release an additional production lot with planned start and due date in that period (PLT = 0). MRP stays near its fixed PLT except when last-minute shortages force an expedited production lot with PLT = 0.

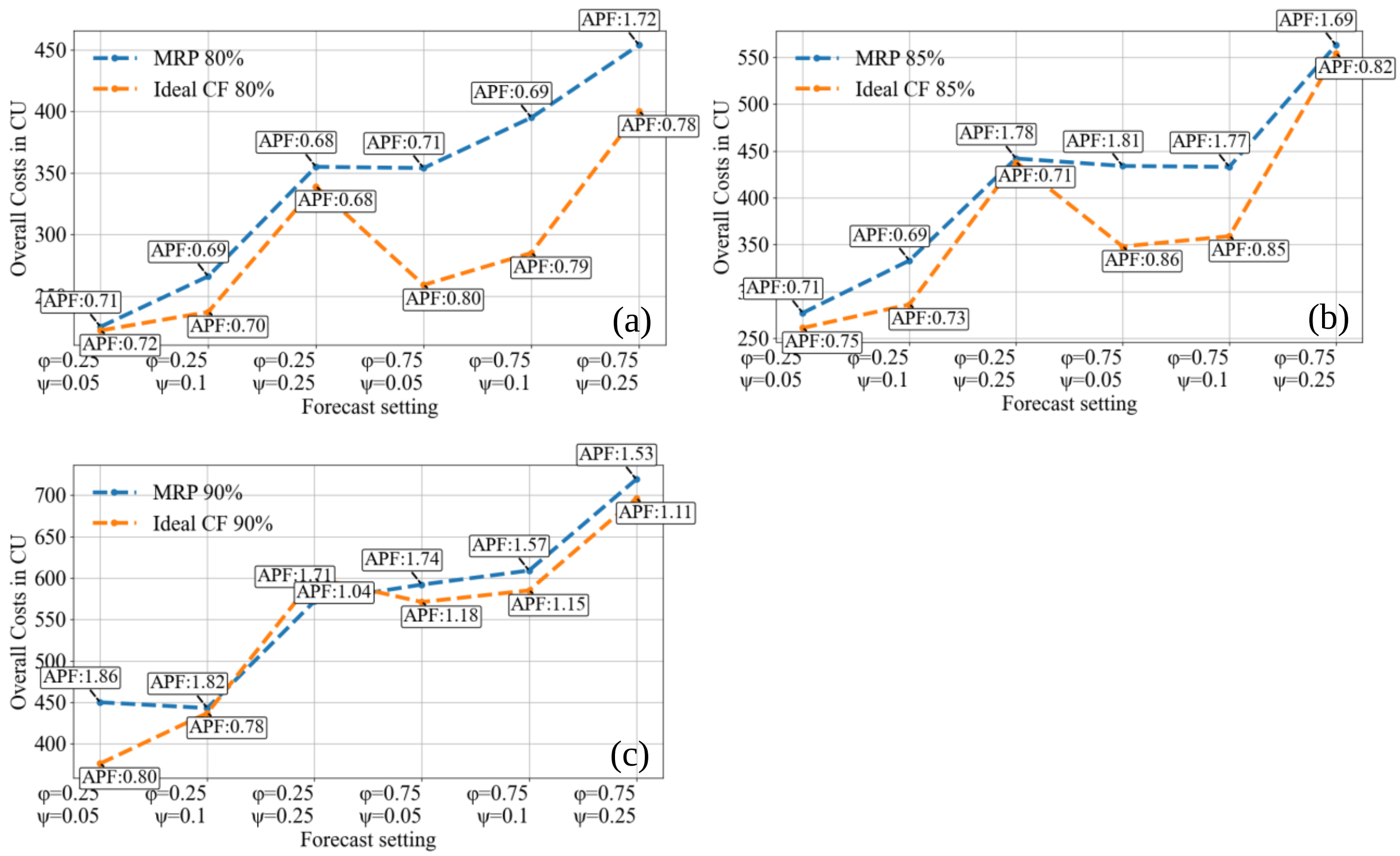

Figure 7: Minimum overall costs for PS1 under imperfect forecast information at (a) 80%, (b) 85%, and (c) 90% utilization

As shown in Figure 7 (a) and Table 5 (a), MRP's APF often falls below its fixed PLT because demand uncertainty triggers zero-lead-time orders. This illustrates the rigidity of fixed lead times: when late requirements arise, MRP must release orders without regard to capacity, leading to higher overall costs. In contrast, CF adapts releases dynamically, resulting in lower costs and higher service levels (e.g., at $\varphi$=0.75, $\psi$=0.05). In the imperfect forecast setting, the second update immediately before delivery increases the need for urgent zero-lead-time orders. This forces CF to issue more such releases and is the main reason for its lower APF compared to the perfect forecast setting. MRP's rigid adherence to PLTs acts as a binding constraint; adaptive strategies relax this constraint and therefore achieve better performance.

At 85% utilization (Figure 7 (b), Table 5 (b)), CF maintains a cost advantage, particularly under low and moderate uncertainty, by flexibly balancing inventory and tardiness. As uncertainty increases, the cost gap narrows and MRP sometimes achieves higher service levels, but CF still ensures improved plans by aligning planned flow times to capacity. This echoes Altendorfer, Felberbauer, and Jodlbauer (2016), who found that forecast errors shift optimal utilization and raise costs, underscoring the value of adaptive release decisions. At 90% utilization (Figure 7 (c), Table 5 (c)), CF again shows more stable performance, reducing costs in most scenarios despite accepting higher tardiness and WIP to keep inventory lower. MRP with PLT = 2, increases inventory to protect service levels. CF's lower APF values reflect its ability to generate capacity-aligned releases, even under high load.

In summary, while the cost gap between CF and MRP narrows under imperfect forecasts, the planning logic differs: MRP relies on fixed PLTs that ignore demand changes, while CF dynamically adapts release timing. This flexibility explains CF's more robust cost–service balance, supporting prior findings that adaptive methods outperform static PLT-based planning under uncertainty (Sun et al. 2012; Altendorfer, Felberbauer, and Jodlbauer 2016). However, the relative improvement from CF diminishes as utilization increases, with smaller cost advantages observed at 90% compared to 80% utilization, since congestion effects begin to dominate. This is again intuitive; at high utilization levels CF's ability to release orders ahead of congested periods is limited by the lack of available capacity in earlier periods.

## 5.3. Effect of three segmented CF on planning performance in PS1

This section examines the performance of the CF model with three segments, explicitly capturing congestion in addition to limited capacity, as shown in Figure 8. The first segment with $\alpha_1 = 1$ and $\beta_1 = 0$, and the third segment with $\alpha_3 = 0$ and $\beta_3 = 1440$ are fixed, as is the slope $\alpha_2 = 0.276$ of the second segment. The values of the second segment intercept $\beta_2$ are varied from 1152 to 288 as seen in Figure 8, representing a more variable production system as discussed in Chapter 3 of Missbauer and Uzsoy (2020).

Table 5: Minimum overall costs (PS1) under imperfect forecast information for (a) 80%, (b) 85%, and (c) 90% utilization

| (a) 80% Utilization | Forecast setting | | | | | | | | | | | |
|---|---|---|---|---|---|---|---|---|---|---|---|---|
| | $\varphi$=0.25 $\psi$=0.05 | | $\varphi$=0.25 $\psi$=0.1 | | $\varphi$=0.25 $\psi$=0.25 | | $\varphi$=0.75 $\psi$=0.05 | | $\varphi$=0.75 $\psi$=0.1 | | $\varphi$=0.75 $\psi$=0.25 | |
| **Planning Type** | **MRP** | **CF** | **MRP** | **CF** | **MRP** | **CF** | **MRP** | **CF** | **MRP** | **CF** | **MRP** | **CF** |
| **Service level** | 93% | 95% | 90% | 91% | 85% | 82% | 77% | 92% | 77% | 88% | 93% | 82% |
| **Safety stock** | 0.2 | 0.2 | 0.4 | 0.2 | 0.6 | 0.4 | 0.2 | 0.2 | 0.4 | 0.2 | 0.4 | 0.4 |
| **FOP** | 1 | 1 | 1 | 1 | 1 | 1 | 1 | 1 | 1 | 1 | 1 | 1 |
| **PLT** | 1 | - | 1 | - | 1 | - | 1 | - | 1 | - | 2 | - |
| **Overallcosts** | 225 | 222 | 266 | 237** | 355 | 339 | 354 | 259** | 395 | 285** | 454 | 400** |
| **Inventory sales** | 37% | 39% | 42% | 36% | 40% | 34% | 20% | 35% | 25% | 32% | 47% | 31% |
| **WIP sales** | 21% | 22% | 18% | 20% | 14% | 15% | 16% | 23% | 14% | 21% | 13% | 14% |
| **Inventory components** | 15% | 16% | 15% | 15% | 12% | 12% | 8% | 14% | 9% | 13% | 21% | 11% |
| **WIP components** | 20% | 20% | 17% | 20% | 12% | 13% | 13% | 20% | 12% | 18% | 10% | 13% |
| **Tardiness** | 7% | 3% | 8% | 9% | 22% | 26% | 43% | 8% | 40% | 16% | 9% | 31% |

| (b) 85% Utilization | Forecast setting | | | | | | | | | | | |
|---|---|---|---|---|---|---|---|---|---|---|---|---|
| | $\varphi$=0.25 $\psi$=0.05 | | $\varphi$=0.25 $\psi$=0.1 | | $\varphi$=0.25 $\psi$=0.25 | | $\varphi$=0.75 $\psi$=0.05 | | $\varphi$=0.75 $\psi$=0.1 | | $\varphi$=0.75 $\psi$=0.25 | |
| **Planning Type** | **MRP** | **CF** | **MRP** | **CF** | **MRP** | **CF** | **MRP** | **CF** | **MRP** | **CF** | **MRP** | **CF** |
| **Service level** | 84% | 90% | 82% | 85% | 91% | 79% | 98% | 87% | 96% | 85% | 93% | 80% |
| **Safety stock** | 0.2 | 0.2 | 0.4 | 0.2 | 0.2 | 0.6 | 0.2 | 0.2 | 0.2 | 0.2 | 0.6 | 0.6 |
| **FOP** | 1 | 1 | 1 | 1 | 1 | 1 | 1 | 1 | 1 | 1 | 1 | 1 |
| **PLT** | 1 | - | 1 | - | 2 | - | 2 | - | 2 | - | 2 | - |
| **Overallcosts** | 277 | 261 | 333 | 286** | 442 | 438 | 434 | 348** | 433 | 359** | 563 | 554 |
| **Inventory sales** | 30% | 34% | 34% | 31% | 44% | 33% | 46% | 28% | 45% | 27% | 42% | 28% |
| **WIP sales** | 20% | 22% | 17% | 20% | 12% | 13% | 16% | 20% | 16% | 20% | 13% | 13% |
| **Inventory components** | 12% | 14% | 12% | 13% | 20% | 12% | 24% | 11% | 23% | 11% | 19% | 10% |
| **WIP components** | 18% | 22% | 15% | 20% | 11% | 13% | 12% | 18% | 12% | 17% | 10% | 11% |
| **Tardiness** | 20% | 8% | 22% | 16% | 13% | 29% | 2% | 23% | 4% | 25% | 16% | 38% |

| (c) 90% Utilization | Forecast setting | | | | | | | | | | | |
|---|---|---|---|---|---|---|---|---|---|---|---|---|
| | $\varphi$=0.25 $\psi$=0.05 | | $\varphi$=0.25 $\psi$=0.1 | | $\varphi$=0.25 $\psi$=0.25 | | $\varphi$=0.75 $\psi$=0.05 | | $\varphi$=0.75 $\psi$=0.1 | | $\varphi$=0.75 $\psi$=0.25 | |
| **Planning Type** | **MRP** | **CF** | **MRP** | **CF** | **MRP** | **CF** | **MRP** | **CF** | **MRP** | **CF** | **MRP** | **CF** |
| **Service level** | 99% | 80% | 97% | 74% | 90% | 92% | 90% | 92% | 95% | 93% | 91% | 90% |
| **Safety stock** | 0.2 | 0.2 | 0.2 | 0.2 | 0.4 | 0.6 | 0.2 | 0.2 | 0.2 | 0.4 | 0.4 | 0.6 |
| **FOP** | 1 | 1 | 1 | 1 | 1 | 2 | 1 | 2 | 2 | 2 | 2 | 2 |
| **PLT** | 2 | - | 2 | - | 2 | - | 2 | - | 2 | - | 2 | - |
| **Overallcosts** | 450 | 376** | 443 | 437 | 572 | 600 | 592 | 571 | 609 | 585 | 719 | 696 |
| **Inventory sales** | 49% | 23% | 47% | 20% | 38% | 43% | 30% | 34% | 41% | 39% | 39% | 37% |
| **WIP sales** | 14% | 18% | 15% | 16% | 13% | 14% | 15% | 17% | 17% | 17% | 15% | 14% |
| **Inventory components** | 25% | 10% | 23% | 8% | 17% | 17% | 17% | 15% | 19% | 17% | 16% | 15% |
| **WIP components** | 12% | 18% | 13% | 16% | 11% | 14% | 11% | 18% | 14% | 16% | 13% | 14% |
| **Tardiness** | 0% | 31% | 2% | 40% | 21% | 12% | 27% | 16% | 9% | 11% | 17% | 20% |

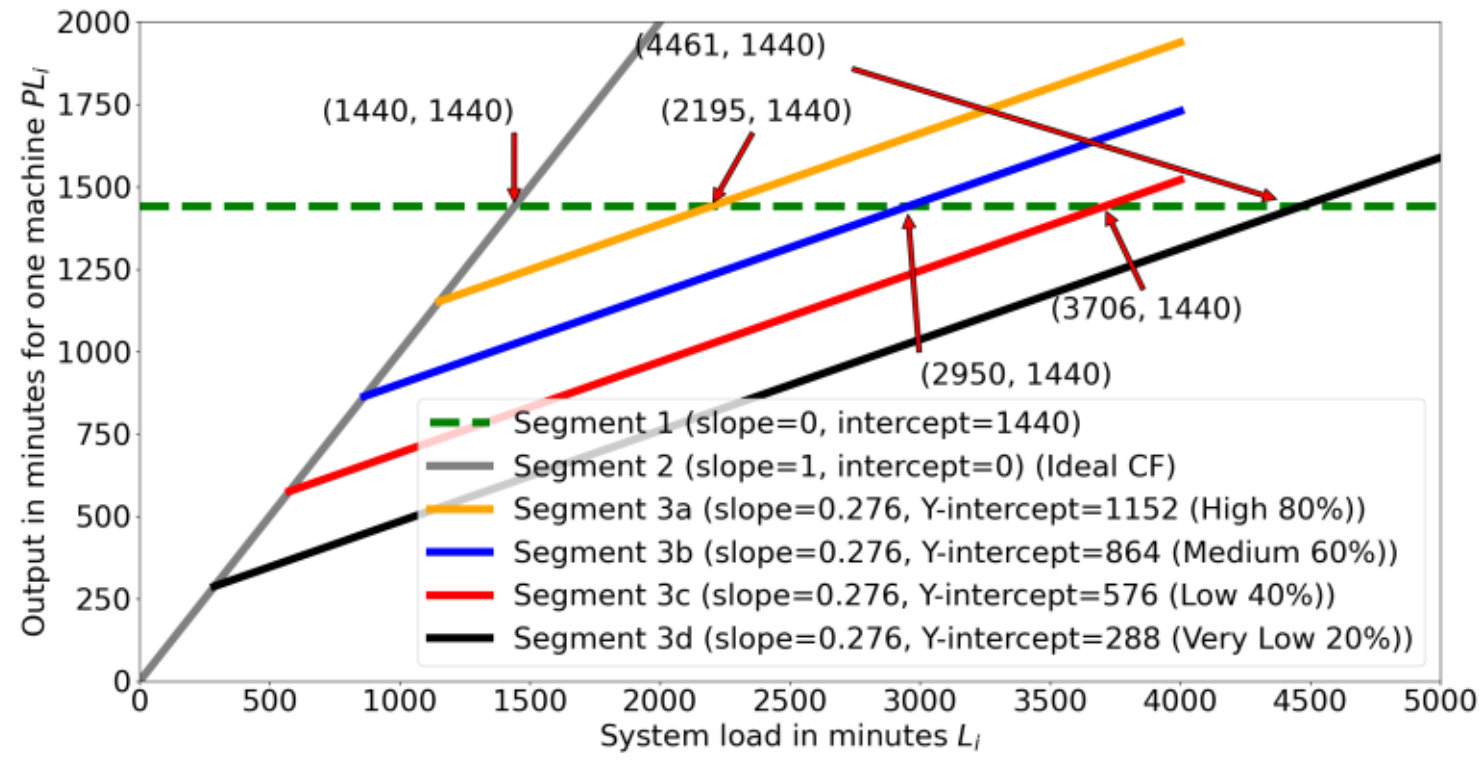

Figure 8: Three-segment clearing functions for PS1

We examine the effect of a third CF segment under 90% utilization with $\varphi$=0.75, $\psi$=0 and $\psi$ =0.05. Figure 9 (a) applies the different CF parameters to the optimal parameter setting from Figure 5 (c) (90% utilization, $\varphi$ =0.75, FOP = 1, SS = 0). As $\beta_2$ increases, representing increased variability in the production process, overall costs rise from 387 CU (Ideal/High) to 898 CU (Very Low), with APF increasing from 1.29 to 4.09. Lower $\beta_2$ levels reduce tardiness costs but rely on early releases, driving up inventory and WIP costs. In the Very Low CF case (20%) the machine's actual average utilization is 90%, almost the same as in the Ideal CF case (89%), but much higher variability due to early releases can be observed. Compared to Ideal CF, periods of full capacity use alternate with phases of underutilization. High 80% achieves the same costs as Ideal CF with less capacity, highlighting the ability of CF to accommodate some variability by early releases. The cost structure shifts notably: inventory components grow from 10% to 20%, and WIP components from 19% to 36%, reflecting earlier lot releases.

Figure 9 (b) shows that under high demand uncertainty, overall costs rise sharply with $\beta_2$ - from 571 CU (Ideal CF) to 1,046 CU (Very Low). This cost increase is driven by longer cycle times (APF 1.18 to 3.87) and higher inventory and WIP costs. Lower CF levels represent more variable production resources, which require earlier releases and higher WIP and inventory to maintain improve service levels and eliminate tardiness.

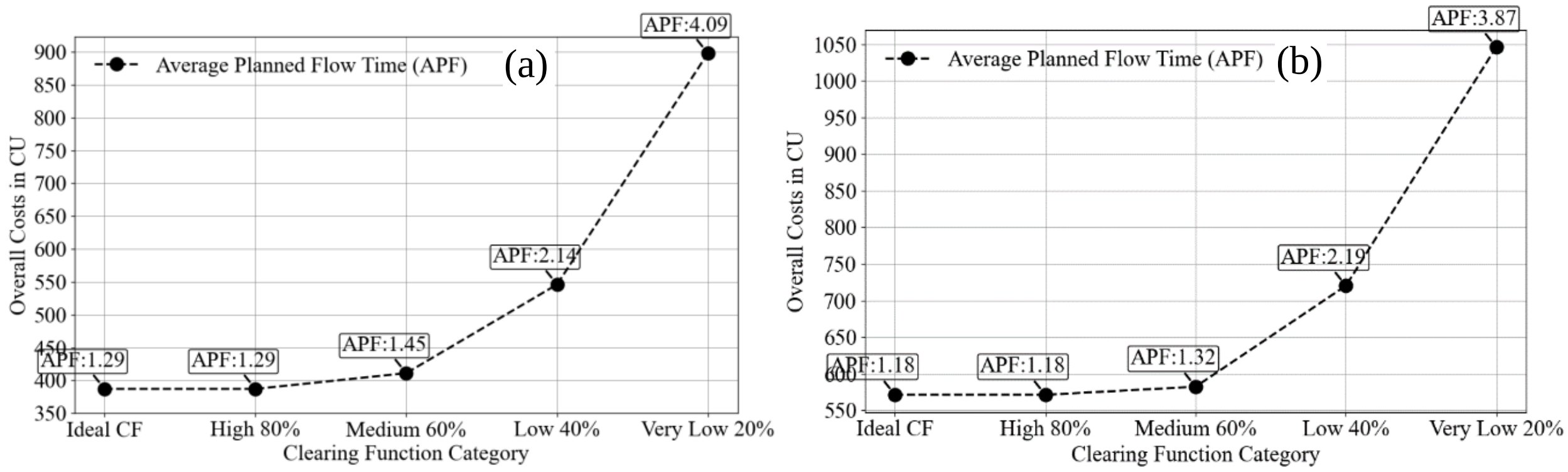

| | Ideal CF | High 80% | Medium 60% | Low 40% | Very Low 20% |
|---|---|---|---|---|---|
| **Service level** | 92% | 92% | 93% | 97% | 98% |
| **Safety stock** | 0 | 0 | 0 | 0 | 0 |
| **FOP** | 1 | 1 | 1 | 1 | 1 |
| **Overallcosts** | 387 | 387 | 411 | 546 | 898 |
| **Inventory sales** | 22% | 22% | 21% | 28% | 23% |
| **WIP sales** | 20% | 20% | 19% | 19% | 16% |
| **Inventory components** | 10% | 10% | 12% | 20% | 19% |
| **WIP components** | 19% | 19% | 20% | 21% | 36% |
| **Tardiness** | 29% | 29% | 28% | 12% | 6% |

| | Ideal CF | High 80% | Medium 60% | Low 40% | Very Low 20% |
|---|---|---|---|---|---|
| **Service level** | 92% | 92% | 91% | 98% | 99% |
| **Safety stock** | 0.2 | 0.2 | 0.1 | 0.1 | 0.1 |
| **FOP** | 2 | 2 | 2 | 2 | 2 |
| **Overallcosts** | 571 | 571 | 582 | 720 | 1046 |
| **Inventory sales** | 33% | 33% | 30% | 36% | 33% |
| **WIP sales** | 17% | 17% | 16% | 18% | 18% |
| **Inventory components** | 15% | 15% | 16% | 19% | 19% |
| **WIP components** | 18% | 18% | 19% | 21% | 27% |
| **Tardiness** | 17% | 17% | 19% | 6% | 3% |

Figure 9: Impact of CF segments on overall costs for **a)** perfect forecast setting $\varphi = 0.75$ and **b)** imperfect forecast setting ($\varphi = 0.75$, $\psi = 0.05$) under, 90% utilization

## 5.4. Complex Production System (PS2)

Production System 2 (PS2), illustrated in Figure 10, represents a multi-item, multi-stage, multi-level flow shop with uniform machine capacities, diverse item–machine assignments, and 32 end items. A stage refers to the set of machines operating at the same BOM level, such as M101–M104 at Level 0 and M201–M204 at Level 1. This system reflects industries such as automotive components and industrial machinery, and provides a more challenging testbed. We target an average utilization of 90% by simulating MRP with FOP = 1 and a daily demand of 47.06 units per end item over 200 days in a system with four machines per stage. Setup times are 10% of machine capacity (144 minutes per period) and vary by BOM level depending on the number of lots processed on a machine (e.g., 18 minutes at level 0, 36 at level 1, 72 at level 2, and 144 at level 3). Uncertainty in setup times is introduced with CV = 0.2. Processing times are deterministic but differ across items within each BOM level, while the same CF parameters are applied uniformly to all items processed on a given machine.

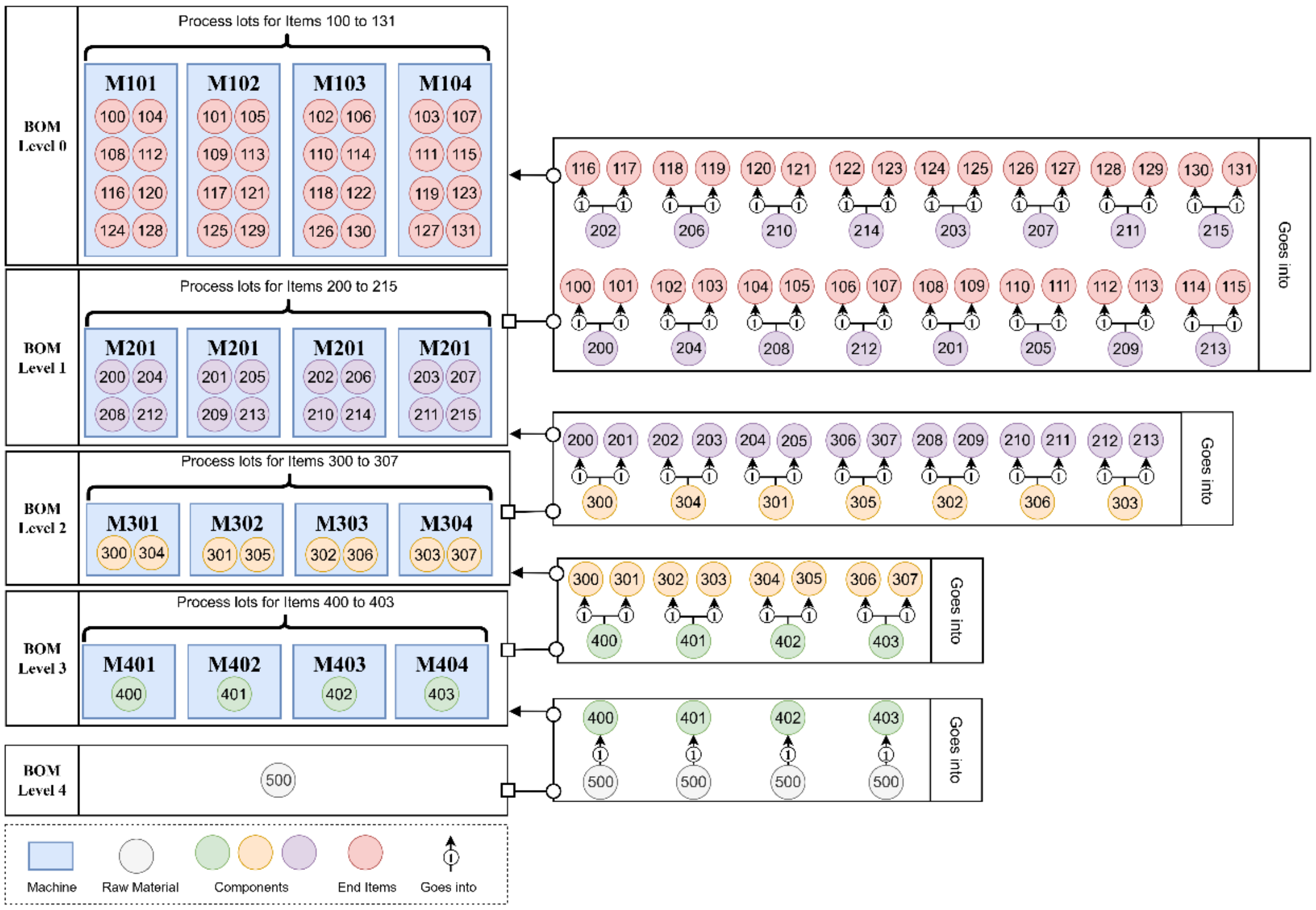

Figure 10: Complex production system 2 (PS2)

## 5.4.1. Experimental design for PS2

To evaluate MRP and CF, the same planning parameters from the simple production system (Table 2) are applied, with two differences. First, the perfect forecast setting is tested across $\varphi = 0$ to 1.0. Due to the complex structure of PS2, the forecast window is set to $H$ =21 periods, as preliminary experiments showed that shorter horizons resulted in planning nervousness causing unstable release patterns in which the relative performance of MRP and CF shifted unpredictably.

## 5.5. Results under Perfect Forecasts (PS2)

This section presents results under perfect forecasts. The cost structure, runtime, and number of replications match PS1, consistent with Tables 2 and 3. As demand variability $\varphi$ increases, CF and MRP show distinct cost and scheduling patterns, as shown in Table 6 and Figure 11.

Table 6: Minimum overall costs for PS2 under perfect forecasts

| 90% Utilization | Forecast setting | | | | | | | | | |
|---|---|---|---|---|---|---|---|---|---|---|
| | $\varphi$=0 | | $\varphi$=0.25 | | $\varphi$=0.5 | | $\varphi$=0.75 | | $\varphi$=1.0 | |
| **Planning Type** | **MRP** | **CF** | **MRP** | **CF** | **MRP** | **CF** | **MRP** | **CF** | **MRP** | **CF** |
| **Service level** | 100% | 100% | 99% | 100% | 95% | 100% | 92% | 99% | 91% | 99% |
| **Safety stock** | 0 | 0 | 0 | 0 | 0 | 0 | 0 | 0 | 0 | 0 |
| **FOP** | 1 | 1 | 1 | 1 | 1 | 1 | 1 | 1 | 1 | 1 |
| **PLT** | 1 | - | 1 | - | 1 | - | 1 | - | 1 | - |
| **Overallcosts** | 2996 | 2996 | 3003** | 3326 | 3072** | 3628 | 3306 | 3701 | 3233** | 3775 |
| **Inventory sales** | 21% | 21% | 21% | 20% | 19% | 19% | 17% | 19% | 17% | 19% |
| **WIP sales** | 12% | 12% | 12% | 11% | 12% | 10% | 12% | 10% | 12% | 10% |
| **Inventory components** | 18% | 18% | 18% | 19% | 16% | 20% | 14% | 21% | 14% | 21% |
| **WIP components** | 49% | 49% | 49% | 50% | 49% | 51% | 46% | 50% | 48% | 50% |
| **Tardiness** | 0% | 0% | 0% | 0% | 4% | 0% | 11% | 0% | 9% | 0% |

At 90% utilization under perfect forecasts (Table 6), MRP maintains lower overall costs than CF because its fixed PLT avoids premature releases and keeps inventory and WIP low. CF, while ensuring near-perfect service ( ≥ 99%), does so by releasing items early, increasing holding costs. This earlier release in CF leads to a cascading effect over multiple BOM levels, i.e. components are released even earlier by CF than MRP, creating unnecessary inventory under perfect forecasts.

To illustrate the cascading effect for CF, we consider the $\varphi$ =0.5 forecast setting for one replication from Table 6. The APF is given by the distribution of durations (Planned End – Planned Start, in periods), which shifts noticeably across BOM levels, compare to Table 7:

Table 7: Planned flow time distribution and APF by BOM level.

| | Planned Flow Time Frequency | | | APF |
|---|---|---|---|---|
| BOM Level | 0 | 1 | 2 | |
| 0 | 0 | 0.950 | 0.050 | 1.05 |
| 1 | 0 | 0.923 | 0.077 | 1.08 |
| 2 | 0.001 | 0.820 | 0.179 | 1.18 |
| 3 | 0.010 | 0.600 | 0.390 | 1.8 |

The results reveal how a narrow planned flow time distribution at higher BOM levels (≈95% one-day durations at Level 0) gradually widens at lower levels. At Level 3, only 60% remain at one day, and nearly 40% extend to two days. This is due to the precedence relations between items in the BOM tree, with higher cumulative lead time to end item at lower BOM levels, requiring components to be completed earlier in the planning horizon. Broadly, CF yields lower tardiness costs with higher inventory levels than MRP for both components and finished inventories. MRP releases orders later, reducing end item inventory levels and incurring some tardiness costs which are offset by the reduced inventory levels. This observation is to be expected in a situation where the CF model operates locally, at a single level of the BOM tree. Detailed results for all perfect forecast settings ($\varphi$ = 0.25, 0.5, 0.75, 1.0) are reported in Table 11 of the Appendix.

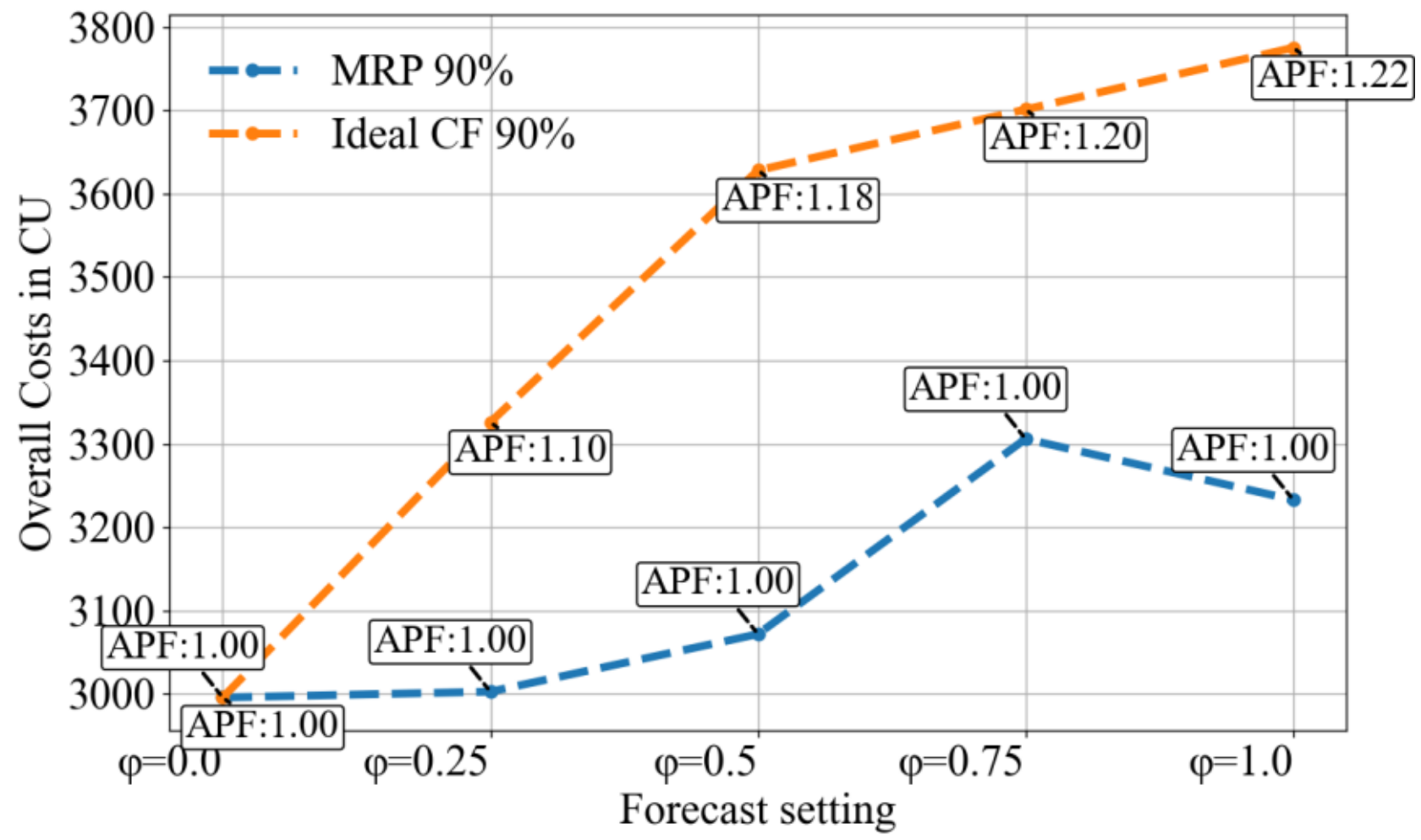

Figure 11: Minimum overall costs for PS2 under 90% utilization and perfect information

## 5.6. Results under Imperfect Forecasts (PS2)

Under imperfect forecasts (Figure 12 and Table 8), CF outperforms MRP by up to 1,500 CU at high uncertainty ($\varphi$= 0.75, $\psi$ = 0.25). This advantage stems from CF's flexible lead times and dynamic release timing, reflected in lower APF values (0.83 vs. MRP's rigid 1.91), which enable CF to align production with capacity and absorb demand updates proactively. In addition, the structure of PS2, with multiple end items sharing common components, creates a risk-pooling effect (Snyder and Shen 2019): aggregated demand at shared components tends to smooth demand fluctuations and reduce variability at that level compared to individual end items. The comparison between MRP and CF across BOM levels confirms the presence of a risk-pooling effect: variability reductions are strongest at the top levels of the BOM, moderate at intermediate levels, and vanish at the lowest levels. While pooling exists under both planning approaches, CF consistently shows stronger reductions, particularly at intermediate levels (see Table 11 in the Appendix for detailed results). CF can potentially benefit from this pooling effect more than MRP, since it allows different PLTs for individual production orders, allowing demand fluctuations to be absorbed at shared resources rather than causing high inventories.

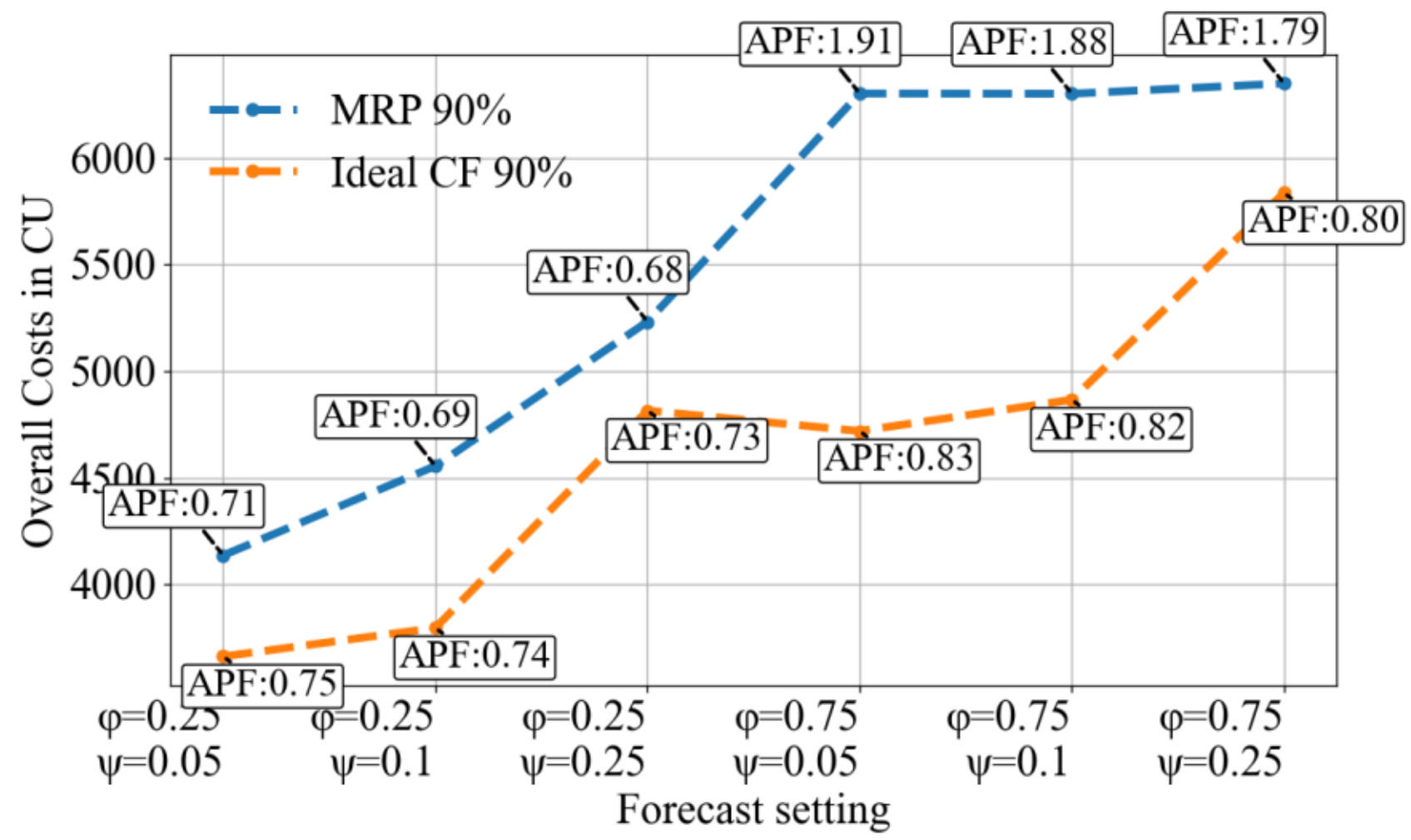

Figure 12: Minimum overall costs - (PS2), 90% utilization, imperfect forecast information

For lower demand variability ($\varphi$ = 0.25, $\psi \in \{0.05,0.1,0.25\}$) MRP's fixed PLTs result in late releases, which are mitigated by higher safety stocks that increase end item inventory costs (up to 40%), while CF maintains end item inventory costs at or below 28% by shifting risk into anticipatory WIP components (up to 48%) with lower safety stocks. For $\varphi$ = 0.75, MRP increases PLT to 2 which increases inventory significantly, permitting a safety stock of 0. Especially in these cases CF outperforms MRP based on the dynamically set lead times. Even when CF accepts higher tardiness costs (19% vs. 8%), it maintains superior performance by adapting release timing, whereas MRP's rigidity leads to excess inventories and longer planned lead times as uncertainty increases. These findings are consistent with prior research: Albey, Yanikoglu, and Uzsoy (2019) show that production planning with fixed PLTs performs poorly under demand uncertainty, while robust adaptive approaches mitigate cost spikes. Similarly, Dolgui, Grimaud, and Shchamialiova (2010) conclude that adaptive rules outperform fixed MRP assumptions.

## 5.7. Impact of Lot Sizing Policy

To assess the impact of an alternative lot-sizing policy, the best parameterizations of PS1 and PS2 (Tables 4,5,6,8) were re-run with fixed order quantities (20, 40, 60, 80, 100 units). For PS1 under perfect forecasts, shown in Table 9, FOQ never outperforms FOP, with a lot size of 20 consistently yielding the lowest costs. FOQ has consistently higher costs because demand forecasts must be rounded up to multiples of 20, creating surplus output and higher holding costs. Limited demand-matching and the propagation of early overproduction through lower levels of the BOM exacerbate this effect. As utilization or demand variability increase, the FOQ–FOP gap narrows especially under CF, where congestion and workload-dependent lead times dominate costs, reducing the relative penalty of lot-size rounding (Jeunet and Jonard 2000; Dolgui and Prodhon 2007). Under MRP, by contrast, fixed PLTs cannot capture congestion effects, so higher utilization or variability do not bring consistent improvements (Jodlbauer and Reitner 2012).

Table 8: Minimum overall costs - (PS2), imperfect forecast information

| **90% Utilization** | **Forecast setting** | | | | | | | | | | | |
|---|---|---|---|---|---|---|---|---|---|---|---|---|
| | **$\varphi$=0.25 $\psi$=0.05** | | **$\varphi$=0.25 $\psi$=0.1** | | **$\varphi$=0.25 $\psi$=0.25** | | **$\varphi$=0.75 $\psi$=0.05** | | **$\varphi$=0.75 $\psi$=0.1** | | **$\varphi$=0.75 $\psi$=0.25** | |
| **Planning Type** | **MRP** | **CF** | **MRP** | **CF** | **MRP** | **CF** | **MRP** | **CF** | **MRP** | **CF** | **MRP** | **CF** |
| **Service level** | 95% | 96% | 93% | 92% | 96% | 89% | 92% | 89% | 89% | 89% | 83% | 86% |
| **Safety stock** | 0.6 | 0.2 | 0.6 | 0.2 | 1.2 | 0.6 | 0 | 0.2 | 0 | 0.4 | 0 | 0.6 |
| **FOP** | 1 | 1 | 1 | 1 | 1 | 1 | 1 | 1 | 1 | 1 | 1 | 1 |
| **PLT** | 1 | - | 1 | - | 1 | - | 2 | - | 2 | - | 2 | - |
| **Overallcosts** | 4136 | 3662** | 4555 | 3796* | 5232 | 4815** | 6302 | 4719** | 6301 | 4866** | 6349 | 5836 |
| **Inventory sales** | 33% | 23% | 29% | 22% | 40% | 28% | 27% | 17% | 26% | 22% | 25% | 23% |
| **WIP sales** | 9% | 11% | 9% | 10% | 8% | 8% | 7% | 9% | 7% | 9% | 7% | 7% |
| **Inventory components** | 19% | 17% | 16% | 17% | 19% | 18% | 36% | 15% | 36% | 18% | 33% | 17% |
| **WIP components** | 38% | 48% | 34% | 46% | 30% | 37% | 26% | 42% | 26% | 41% | 27% | 34% |
| **Tardiness** | 1% | 1% | 12% | 5% | 3% | 9% | 4% | 17% | 5% | 10% | 8% | 19% |

Table 9: Relative percentage cost increase of FOQ compared to FOP under perfect forecasts.

| | | **Forecast setting** | | | | | | | | | | | | | |
|---|---|---|---|---|---|---|---|---|---|---|---|---|---|---|---|
| | | **$\varphi$=0** | | **$\varphi$=0.25** | | **$\varphi$=0.5** | | **$\varphi$=0.75** | | **$\varphi$=1.0** | | **$\varphi$=1.25** | | **$\varphi$=1.5** | |
| **Prod. System** | **Utilization** | **MRP** | **CF** | **MRP** | **CF** | **MRP** | **CF** | **MRP** | **CF** | **MRP** | **CF** | **MRP** | **CF** | **MRP** | **CF** |
| **PS1** | 80% | 24.16% | 23.60% | 23.50% | 22.40% | 20.26% | 19.21% | 17.52% | 18.52% | 17.65% | 18.06% | 16.26% | 17.97% | 14.80% | 16.89% |
| | 85% | 37.77% | 32.98% | 27.40% | 21.46% | 18.15% | 16.88% | 8.45% | 12.83% | 8.09% | 13.57% | 8.09% | 14.96% | 7.35% | 13.75% |
| | 90% | 57.58% | 45.96% | 23.72% | 17.65% | 7.21% | 13.29% | 5.14% | 5.94% | 7.22% | 7.93% | 6.80% | 5.21% | 4.65% | 4.20% |
| **PS2** | 90% | 11.98% | 23.73% | 26.84% | 18.31% | 13.87% | 10.23% | 95.46% | 9.73% | 7.42% | 7.66% | | | | |

Considering imperfect forecast information in Table 10, the trend of decreasing relative differences between FOQ and FOP costs with increasing demand fluctuations and utilization becomes more pronounced, occasionally resulting in FOQ outperforming FOP (green cells). The lowest FOQ cost is not always achieved with lot size of 20 units: under MRP, in 16 cases a larger FOQ yields lower costs, whereas for CF only two such cases occur. FOQ outperforms FOP in 14 cases for CF, with cost reductions of up to 19.66%; for MRP this occurs only twice, with a maximum reduction of 2%.

Table 10: Relative percentage increase of FOQ compared to FOP under imperfect forecasts.

| *FOQ cost reduction* ↓ | | | **Forecast setting** | | | | | | | | | | | |
|---|---|---|---|---|---|---|---|---|---|---|---|---|---|---|
| *FOQ cost increase* ↑ | | | $\varphi$=0.25 $\psi$=0.05 | | $\varphi$=0.25 $\psi$=0.1 | | $\varphi$=0.25 $\psi$=0.25 | | $\varphi$=0.75 $\psi$=0.05 | | $\varphi$=0.75 $\psi$=0.1 | | $\varphi$=0.75 $\psi$=0.25 | |
| **Prod. System** | **Utilization** | | **MRP** | **CF** | **MRP** | **CF** | **MRP** | **CF** | **MRP** | **CF** | **MRP** | **CF** | **MRP** | **CF** |
| **PS1** | **80%** | Diff. | 87.11% | 22.97% | 63.53% | 16.88% | 37.18% | 0.88% | 35.59% | 14.29% | 28.61% | 7.37% | 28.41% | -7.62% |
| | | FOP | 1 | 1 | 1 | 1 | 1 | 1 | 1 | 1 | 1 | 1 | 1 | 1 |
| | | FOQ | 60 | 20 | 60 | 20 | 60 | 20 | 60 | 20 | 60 | 20 | 60 | 20 |
| | **85%** | Diff. | 71.12% | 15.71% | 59.76% | 9.44% | 26.70% | -3.20% | 16.36% | -2.30% | 31.87% | -4.18% | 17.05% | -11.91% |
| | | FOP | 1 | 1 | 1 | 1 | 1 | 1 | 1 | 1 | 1 | 1 | 1 | 1 |
| | | FOQ | 60 | 20 | 60 | 20 | 60 | 20 | 40 | 20 | 60 | 20 | 60 | 20 |
| | **90%** | Diff. | 12.44% | -8.24% | 34.76% | -17.39% | 7.52% | -13.17% | -1.52% | -18.39% | 2.79% | -19.66% | 2.36% | -11.62% |
| | | FOP | 1 | 1 | 1 | 1 | 1 | 2 | 1 | 2 | 2 | 2 | 2 | 2 |
| | | FOQ | 20 | 20 | 60 | 20 | 60 | 40 | 40 | 20 | 60 | 20 | 80 | 60 |
| **PS2** | **90%** | Diff. | 380.44% | 13.52% | 412.18% | 9.62% | 213.51% | 3.61% | -2.00% | -4.98% | 25.47% | -1.69% | 88.47% | -5.23% |
| | | FOP | 1 | 1 | 1 | 1 | 1 | 1 | 1 | 1 | 1 | 1 | 1 | 1 |
| | | FOQ | 20 | 20 | 20 | 20 | 20 | 20 | 20 | 20 | 20 | 20 | 20 | 20 |

Even though the general trend toward lower FOQ costs with increasing uncertainty and utilization can be investigated, the performance of MRP for PS2 between FOQ and FOP in the forecast settings $\varphi = 0.25$, $\psi \in \{0.05/0.1/0.25\}$ is very poor. This is mainly due to tardiness; e.g., for ($\varphi = 0.25$, $\psi = 0.05$) we see an absolute increase from 62 CU to 16,528 CU, for ($\varphi = 0.25$, $\psi = 0.1$) from 528 CU up to 20,061 CU, and for ($\varphi = 0.25$, $\psi = 0.25$) from 570 CU to 7,172 CU. An FOQ of 20 units provides the closest performance to FOP = 1. Because SS and PLT are fixed in the tested scenarios, FOQ would require a higher PLT and/or SS to obtain lower costs in MRP.

By contrast, CF does not show extreme cost increases with FOQ. Even when only FOQ is varied, lead times adjust endogenously with workload, mitigating the mismatch between order frequency and demand. As a result, tardiness remains low, and the FOQ–FOP differences are moderate. This highlights CF's additional degree of freedom (production order based lead time) and its greater flexibility in parameter tuning, which is decisive in PS2's multi-stage setting. Thus CF benefits substantially more from FOQ than MRP.

# 6. Conclusion

We present an exploratory comparison of traditional MRP and CF-based release planning in multi-item, multi-stage production systems with rolling-horizon updates and stochastic demand. The results demonstrate considerable potential for improving traditional MRP performance by replacing the backward scheduling step with a CF-based optimization model, especially when demand uncertainty is high. The numerical study further clarifies the conditions under which these benefits arise: (i) CF performs best when uncertainty is substantial, demand updates occur close to delivery, or utilization is high, as workload-dependent lead times enable proactive congestion management. (ii) When variability and congestion pressures are low, fixed PLTs can provide effective release timing, limiting the relative advantage of CF. (iii) In multi-level production systems, shared resources create risk-pooling effects that CF can exploit to absorb variability. (iv) CF also adapts well to different lot-sizing policies; even under rigid FOQ settings—where MRP often becomes costly—its order-specific timing better accommodates fixed batch sizes.

Taken together, these insights show that a naive CF integration can improve performance in stochastic, capacity-constrained, and diverging production systems, but only within clear limits. The approach relies on several restrictive assumptions: (a) items sharing a resource must lie within the same BOM level; (b) MRP's lot sizes and release sequences are preserved; and (c) CF is applied locally at each BOM level without modeling cross-level capacity interactions. These constraints mean the method is not intended as a universal replacement for backward scheduling. Instead, it serves as a modular enhancement for environments with high variability and tight capacity, where workload-dependent, order-specific release timing provides the greatest benefit.

A number of future directions can be considered. The first of these is to relax the assumption that all items sharing a production resource are at the same level of the BOM tree. This is often not the case in practice, but brings the need to consider multiple BOM levels simultaneously in the release planning models, which leads to significantly more complicated models. Further computational experiments with the proposed approach examining the impact of more complex lot-sizing procedures (e.g., Wagner-Whitin) as well as different BOM and cost structures are also required to identify opportunities for further improvements. While many of the limitations of the current approach relative to obtaining optimal solutions arise from the need to maintain consistency with MRP logic, the ability to improve MRP by replacing specific components in a modular manner might well enhance the implementability of the proposed improvements.

## ACKNOWLEDGEMENTS

This research was funded in whole, or in part, by the Austrian Science Fund (FWF) [10.55776/P32954].  The authors declare no competing interests.

## DATA AVAILABILITY STATEMENT

The data that support the findings of this study are available in Zenodo at 10.5281/zenodo.17433801 upon reasonable request to Wolfgang Seiringer.

## APPENDIX

**Example (Forecast Settings Illustration)**

For example, for the simple production system (PS1) and 90% utilization item $g$=100 and $x_{100} = 47.06$, with $H = 10$ and $\gamma = 11 \rightarrow 0$ and due date $d$=20:

*Perfect forecast settings*:
$(\varphi, \psi) = (0{,}0) \rightarrow D_{100\ 20\ 11\ldots0} = (47, \ldots, 47)$
$(\varphi, \psi) = (0.25{,}0) \rightarrow D_{100\ 20\ 11\ldots0} = (47{,}43 \ldots, 43)$
*Imperfect forecast*:
$(\varphi, \psi) = (0.25{,}0.25) \rightarrow D_{100\ 20\ 11\ldots0} = (47{,}25, \ldots, 25{,}28)$

Table 11: Risk pooling results for MRP and CF across BOM levels in PS2 at 90% utilization.

| | | $\varphi=0.25$ $\psi=0.05$ | | | | | | $\varphi=0.25$ $\psi=0.1$ | | | | | | $\varphi=0.25$ $\psi=0.25$ | | | | | |
|---|---|---|---|---|---|---|---|---|---|---|---|---|---|---|---|---|---|---|---|
| | | *Job count* | | *CV Pooled* | | *CV Reduction* | | *Job count* | | *CV Pooled* | | *CV Reduction* | | *Job count* | | *CV Pooled* | | *CV Reduction* | |
| **BOM level** | **Machine** | **MRP** | **CF** | **MRP** | **CF** | **MRP** | **CF** | **MRP** | **CF** | **MRP** | **CF** | **MRP** | **CF** | **MRP** | **CF** | **MRP** | **CF** | **MRP** | **CF** |
| 0 | M11 | 1808 | 1805 | 0.08 | 0.08 | 68% | 71% | 1864 | 1862 | 0.08 | 0.07 | 70% | 73% | 1875 | 1869 | 0.09 | 0.09 | 76% | 78% |
| | M12 | 1819 | 1819 | 0.08 | 0.09 | 69% | 69% | 1874 | 1873 | 0.08 | 0.09 | 72% | 71% | 1891 | 1889 | 0.10 | 0.09 | 76% | 77% |
| | M13 | 1860 | 1858 | 0.09 | 0.08 | 67% | 70% | 1890 | 1889 | 0.08 | 0.09 | 71% | 71% | 1933 | 1931 | 0.09 | 0.09 | 80% | 80% |
| | M14 | 1822 | 1824 | 0.08 | 0.08 | 70% | 73% | 1884 | 1882 | 0.08 | 0.09 | 73% | 73% | 1894 | 1863 | 0.10 | 0.09 | 77% | 79% |
| 1 | M21 | 916 | 934 | 0.11 | 0.12 | 47% | 60% | 939 | 961 | 0.11 | 0.12 | 53% | 62% | 927 | 969 | 0.14 | 0.14 | 63% | 70% |
| | M22 | 903 | 933 | 0.09 | 0.09 | 56% | 61% | 919 | 959 | 0.09 | 0.10 | 59% | 66% | 933 | 923 | 0.11 | 0.15 | 68% | 63% |
| | M23 | 894 | 925 | 0.12 | 0.11 | 51% | 54% | 907 | 943 | 0.13 | 0.10 | 50% | 61% | 939 | 944 | 0.12 | 0.11 | 61% | 66% |
| | M24 | 901 | 921 | 0.12 | 0.11 | 46% | 63% | 917 | 937 | 0.10 | 0.13 | 53% | 63% | 947 | 917 | 0.11 | 0.13 | 65% | 70% |
| 2 | M31 | 452 | 450 | 0.29 | 0.37 | 27% | 29% | 461 | 459 | 0.27 | 0.39 | 27% | 29% | 462 | 468 | 0.28 | 0.31 | 15% | 42% |
| | M32 | 445 | 470 | 0.29 | 0.37 | 26% | 20% | 458 | 474 | 0.29 | 0.37 | 23% | 25% | 478 | 483 | 0.28 | 0.29 | 32% | 35% |
| | M33 | 456 | 466 | 0.27 | 0.25 | 23% | 28% | 476 | 466 | 0.25 | 0.25 | 26% | 31% | 456 | 469 | 0.27 | 0.23 | 21% | 42% |
| | M34 | 492 | 496 | 0.18 | 0.19 | 36% | 34% | 506 | 501 | 0.15 | 0.19 | 40% | 42% | 481 | 461 | 0.17 | 0.24 | 39% | 44% |
| 3 | M41 | 231 | 230 | 0.46 | 0.57 | 0% | 0% | 229 | 224 | 0.46 | 0.58 | 0% | 0% | 226 | 214 | 0.43 | 0.53 | 0% | 0% |
| | M42 | 252 | 243 | 0.40 | 0.52 | 0% | 0% | 255 | 248 | 0.30 | 0.43 | 0% | 0% | 239 | 237 | 0.34 | 0.42 | 0% | 0% |
| | M43 | 219 | 226 | 0.43 | 0.52 | 0% | 0% | 228 | 231 | 0.40 | 0.52 | 0% | 0% | 246 | 244 | 0.38 | 0.40 | 0% | 0% |
| | M44 | 243 | 243 | 0.36 | 0.47 | 0% | 0% | 245 | 244 | 0.30 | 0.46 | 0% | 0% | 231 | 236 | 0.43 | 0.41 | 0% | 0% |

| | | $\varphi=0.75$ $\psi=0.05$ | | | | | | $\varphi=0.75$ $\psi=0.1$ | | | | | | $\varphi=0.75$ $\psi=0.25$ | | | | | |
|---|---|---|---|---|---|---|---|---|---|---|---|---|---|---|---|---|---|---|---|
| | | *Job count* | | *CV Pooled* | | *CV Reduction* | | *Job count* | | *CV Pooled* | | *CV Reduction* | | *Job count* | | *CV Pooled* | | *CV Reduction* | |
| **BOM level** | **Machine** | **MRP** | **CF** | **MRP** | **CF** | **MRP** | **CF** | **MRP** | **CF** | **MRP** | **CF** | **MRP** | **CF** | **MRP** | **CF** | **MRP** | **CF** | **MRP** | **CF** |
| 0 | M11 | 1358 | 1807 | 0.16 | 0.13 | 67% | 75% | 1404 | 1828 | 0.16 | 0.12 | 70% | 76% | 1483 | 1820 | 0.15 | 0.12 | 73% | 79% |
| | M12 | 1380 | 1841 | 0.15 | 0.12 | 72% | 77% | 1401 | 1846 | 0.15 | 0.11 | 72% | 79% | 1499 | 1806 | 0.14 | 0.10 | 75% | 83% |
| | M13 | 1400 | 1791 | 0.13 | 0.11 | 74% | 81% | 1414 | 1851 | 0.16 | 0.13 | 70% | 77% | 1487 | 1838 | 0.15 | 0.12 | 73% | 80% |
| | M14 | 1388 | 1789 | 0.15 | 0.10 | 72% | 81% | 1415 | 1833 | 0.17 | 0.12 | 67% | 77% | 1521 | 1770 | 0.15 | 0.12 | 73% | 81% |
| 1 | M21 | 667 | 939 | 0.16 | 0.14 | 64% | 74% | 670 | 897 | 0.17 | 0.17 | 61% | 68% | 708 | 913 | 0.16 | 0.15 | 64% | 72% |
| | M22 | 660 | 917 | 0.14 | 0.15 | 66% | 68% | 676 | 955 | 0.16 | 0.16 | 63% | 65% | 712 | 943 | 0.18 | 0.15 | 61% | 67% |
| | M23 | 680 | 924 | 0.17 | 0.20 | 60% | 63% | 678 | 940 | 0.18 | 0.15 | 60% | 68% | 733 | 928 | 0.18 | 0.16 | 60% | 75% |
| | M24 | 668 | 883 | 0.16 | 0.16 | 64% | 66% | 682 | 923 | 0.16 | 0.14 | 62% | 69% | 730 | 868 | 0.19 | 0.19 | 62% | 69% |
| 2 | M31 | 330 | 473 | 0.34 | 0.37 | 13% | 43% | 332 | 472 | 0.31 | 0.37 | 20% | 43% | 347 | 456 | 0.30 | 0.37 | 29% | 42% |
| | M32 | 328 | 452 | 0.35 | 0.35 | 22% | 40% | 328 | 467 | 0.41 | 0.38 | 9% | 35% | 344 | 451 | 0.34 | 0.35 | 33% | 43% |
| | M33 | 329 | 462 | 0.39 | 0.41 | 8% | 25% | 335 | 460 | 0.43 | 0.40 | 8% | 29% | 361 | 465 | 0.40 | 0.35 | 18% | 38% |
| | M34 | 343 | 447 | 0.30 | 0.28 | 34% | 43% | 342 | 470 | 0.34 | 0.28 | 33% | 45% | 365 | 455 | 0.31 | 0.28 | 32% | 52% |
| 3 | M41 | 163 | 235 | 0.55 | 0.59 | 0% | 0% | 163 | 229 | 0.59 | 0.63 | 0% | 0% | 171 | 222 | 0.47 | 0.56 | 0% | 0% |
| | M42 | 161 | 225 | 0.57 | 0.56 | 0% | 0% | 167 | 237 | 0.56 | 0.55 | 0% | 0% | 176 | 232 | 0.52 | 0.54 | 0% | 0% |
| | M43 | 172 | 230 | 0.46 | 0.52 | 0% | 0% | 169 | 234 | 0.54 | 0.54 | 0% | 0% | 179 | 227 | 0.40 | 0.50 | 0% | 0% |
| | M44 | 173 | 236 | 0.52 | 0.55 | 0% | 0% | 180 | 234 | 0.50 | 0.50 | 0% | 0% | 190 | 228 | 0.41 | 0.50 | 0% | 0% |

The values in Table 11 are based on daily workload time series per machine, using data after the 40-period warm-up and one replication of the imperfect forecast settings for PS2 reported in Table 8. To construct these series, the simulation time was discretized into daily buckets, and for each machine the total processing time per bucket was recorded as workload. On this basis, we computed the coefficient of variation (CV) for the workload of individual item streams as well as for the aggregated workload of all items processed on the same machine. The risk pooling effect is expressed as the percentage reduction in variability when moving from the average item-level CV to the pooled CV. A high reduction indicates that the aggregation of item streams smooths workload fluctuations, whereas a reduction close to zero implies that no pooling benefit is achieved, for instance when a machine is dedicated to a single item (e.g., machines M41 to M44 at BOM level 3).

Table 12: Average Planning Factor (APF) across BOM levels for perfect forecast settings (***φ***) and complex production system (PS2). LT columns show share of production orders per lead time.

| | Forecast setting | | | | | | | | | | | | | | | | |
|---|---|---|---|---|---|---|---|---|---|---|---|---|---|---|---|---|---|
| | φ=0.25 | | | | φ=0.5 | | | | φ=0.75 | | | | φ=1.0 | | | | |
| **BOM level** | **APF** | **LT=0** | **LT=1** | **LT=2** | **APF** | **LT=0** | **LT=1** | **LT=2** | **APF** | **LT=0** | **LT=1** | **LT=2** | **APF** | **LT=0** | **LT=1** | **LT=2** | **LT=3** |
| 0 | 1.01 | 0.00% | 98.90% | 1.10% | 1.05 | 0.00% | 95.00% | 5.00% | 1.07 | 0.00% | 93.00% | 7.00% | 1.09 | 0.00% | 90.80% | 9.20% | 0.00% |
| 1 | 1.02 | 0.00% | 97.60% | 2.40% | 1.08 | 0.00% | 92.30% | 7.70% | 1.12 | 0.00% | 87.80% | 12.20% | 1.12 | 0.10% | 87.30% | 12.60% | 0.00% |
| 2 | 1.09 | 0.10% | 91.10% | 8.80% | 1.18 | 0.10% | 82.00% | 17.90% | 1.23 | 0.00% | 77.30% | 22.70% | 1.23 | 0.20% | 77.10% | 22.70% | 0.10% |
| 3 | 1.25 | 0.00% | 75.20% | 24.80% | 1.38 | 1.00% | 60.00% | 39.00% | 1.45 | 0.80% | 53.00% | 46.20% | 1.52 | 1.40% | 53.00% | 53.30% | 0.00% |

The Table 12 reports APF values for BOM levels 0–3 under perfect forecast settings ($\varphi$ = 0.25, 0.5, 0.75, 1.0) of the first simulation replication and 90% utilization. LT columns indicate the proportion of production orders assigned to each lead time. Higher BOM levels and lower $\varphi$ values result in larger APF and a broader LT distribution.